\documentclass[structabstract]{aa}  

\usepackage{graphicx}
\usepackage{txfonts}
\usepackage[percent]{overpic}
\usepackage{color}
\usepackage{euscript}
\usepackage{multirow}
\usepackage{upgreek} 
\usepackage{natbib,twoopt} 
\usepackage[breaklinks=true]{hyperref} 
\bibpunct{(}{)}{;}{a}{}{,} 
\usepackage[breaklinks=true]{hyperref}	
\hypersetup{
    bookmarks=true,         
    unicode=false,          
    pdftoolbar=true,        
    pdfmenubar=true,        
    pdffitwindow=false,     
    pdfstartview={FitH},    
    pdftitle={Magnetic fields of opposite polarity in sunspot penumbrae},    
    pdfauthor={Morten Franz},     
    pdfsubject={A&A Paper},   
    pdfcreator={Morten Franz},   
    pdfproducer={Morten Franz}, 
    pdfkeywords={sunspots} {infrared} {photosphere} {surfacemagnetism} {magnetic topology} {penumbra} {spectropolarimetry} {HINODE}, 
    pdfnewwindow=false,      
    colorlinks=true,       
    linkcolor=black,          
    citecolor=blue,        
    filecolor=blue,      
    urlcolor=blue,           
    linkbordercolor={1 0 0}, 	
    citebordercolor={0 1 0}, 
    urlbordercolor={1 1 1},
    linktocpage=false,
    raiselinks=true,
    hyperfootnotes=false
}

 \bibpunct{(}{)}{;}{a}{}{,} 
 \newcommandtwoopt{\citeads}[3][][]{\href{http://adsabs.harvard.edu/abs/#3}{\citealp[#1][#2]{#3}}} 
 \newcommandtwoopt{\citepads}[3][][]{\href{http://adsabs.harvard.edu/abs/#3}{\citep[#1][#2]{#3}}} 
 \newcommandtwoopt{\citetads}[3][][]{\href{http://adsabs.harvard.edu/abs/#3}{\citet[#1][#2]{#3}}}
 \newcommandtwoopt{\citeyearads}[3][][]{\href{http://adsabs.harvard.edu/abs/#3}{\citeyear[#1][#2]{#3}}}

\begin{document}
   \title{Magnetic fields of opposite polarity in sunspot penumbrae}

\author{M. Franz\inst{\ref{inst1}} \and M. Collados\inst{\ref{inst2},\ref{inst3}} \and C. Bethge\inst{\ref{inst1},\ref{inst9}} \and R. Schlichenmaier\inst{\ref{inst1}} \and J.M. Borrero\inst{\ref{inst1}} \and W. Schmidt\inst{\ref{inst1}} \and A.~Lagg\inst{\ref{inst5}} \and S.~K.~Solanki\inst{\ref{inst5},\ref{inst8}} \and T.~Berkefeld\inst{\ref{inst1}} \and C. Kiess\inst{\ref{inst1}} \and R.~Rezaei\inst{\ref{inst1},\ref{inst2},\ref{inst3}} \and D.~Schmidt\inst{\ref{inst1}}  \and M. Sigwarth\inst{\ref{inst1}} \and D.~Soltau\inst{\ref{inst1}} \and R. Volkmer\inst{\ref{inst1}} \and O. von der Luhe\inst{\ref{inst1}} \and T.~Waldmann\inst{\ref{inst1}} \and D.~Orozco\inst{\ref{inst2},\ref{inst3}} \and A.~Pastor~Yabar\inst{\ref{inst2},\ref{inst3}} \and C. Denker\inst{\ref{inst4}} \and H. Balthasar\inst{\ref{inst4}} \and J. Staude\inst{\ref{inst4}} \and A. Hofmann\inst{\ref{inst4}} \and K. Strassmeier\inst{\ref{inst4}} \and A. Feller\inst{\ref{inst5}} \and H. Nicklas\inst{\ref{inst6}} \and F. Kneer\inst{\ref{inst6}} \and M. Sobotka\inst{\ref{inst7}}  }
    \institute{
    	Kiepenheuer Institut f\"ur Sonnenphysik, Sch\"oneckstr. 6, 79104 Freiburg, Germany\label{inst1} \and
    	Instituto de Astrof\'isica de Canarias, c/V\'ia L\'actea s/n, 38205 La Laguna, Spain\label{inst2} \and
	Departamento de Astrof\'isica, Universidad de La Laguna, 38205 La Laguna (Tenerife), Spain \label{inst3} \and
     	Leibniz-Institut f\"ur Astrophysik Potsdam (AIP), An der Sternwarte 16, 14482 Potsdam, Germany\label{inst4} \and
     	Max-Planck-Institut f\"ur Sonnensystemforschung, Justus von Liebig Weg 3, 37077 G\"ottingen, Germany\label{inst5} \and
	Institut f\"ur Astrophysik, Friedrich Hund Platz 1, 37077 G\"ottingen, Germany\label{inst6} \and
     	Astronomical Institute, Academy of Sciences of the Czech Republic, Fri\u{c}ova 298, 25165 Ond\u{r}ejov, Czech Republic\label{inst7} \and
	Kyung Hee University, Yongin, Gyeonggi-Do, 446 701 Republic of Korea\label{inst8} \and
	USRA Huntsville, 6767 Madison Pike \# 450, Huntsville, AL 35806, United States\label{inst9}
	}
   \date{Received 01.03.2016 / Accepted 01.08.2016 / Published ??.??.????}

  \abstract
{A significant part of the penumbral magnetic field returns below the surface in the very deep photosphere. For lines in the visible, a large portion of this return field can only be detected indirectly by studying its imprints on strongly asymmetric and three-lobed Stokes V profiles. Infrared lines probe a narrow layer in the very deep photosphere, providing the possibility of directly measuring the orientation of magnetic fields close to the solar surface. }
{We study the topology of the penumbral magnetic field in the lower photosphere, focusing on regions where it returns below the surface.}
{We analyzed 71 spectropolarimetric datasets from Hinode and from the GREGOR infrared spectrograph. We inferred the quality and polarimetric accuracy of the infrared data after applying several reduction steps. Techniques of spectral inversion and forward synthesis were used to test the detection algorithm. We compared the morphology and the fractional penumbral area covered by reversed-polarity and three-lobed Stokes V profiles for sunspots at disk center. We determined the amount of reversed-polarity and three-lobed Stokes V profiles in visible and infrared data of sunspots at various heliocentric angles. From the results, we computed center-to-limb variation curves, which were interpreted in the context of existing penumbral models.}
{Observations in visible and near-infrared spectral lines yield a significant difference in the penumbral area covered by magnetic fields of opposite polarity. In the infrared, the number of reversed-polarity Stokes V profiles is smaller by a factor of two than in the visible. For three-lobed Stokes V profiles the numbers differ by up to an order of magnitude.}
{}
  \keywords{Sunspots -- Sun: infrared -- Sun: photosphere -- Sun: surface magnetism -- Sun: magnetic topology}

  \maketitle
  \titlerunning{Magnetic fields of opposite polarity in sunspot penumbrae} 
  \authorrunning{Franz et al.}
  

\section{Introduction}
\label{sec:int}

Our knowledge of the way magnetic and velocity fields are configured in the penumbra has greatly improved during the past decade (see, e.g.,
\citet{2003A&ARv..11..153S, 
2011LRSP....8....3R, 
2011LRSP....8....4B} 
for reviews).

One example of this improvement concerns the disappearance of the Evershed flow at the white-light boundary of the sunspots, which seemed to be at odds with the concept of mass conservation. A possible solution, that a significant part of the plasma flow continues into the sunspot canopy, was discarded by observations
\citep{1994A&A...283..221S}. 
The contradiction was solved after spectropolarimetric observation of high quality became available, which allowed for the detection of strong downflow plumes in the outer parts of the penumbra and showed where the plasma of the Evershed flow returns into the Sun 
\citep{1997Natur.389...47W, 
2001ApJ...549L.139D, 
2004A&A...427..319B, 
2009A&A...508.1453F}. 

Another example is the interplay of penumbral magnetic and velocity fields. As these are well aligned 
\citep{2003A&A...403L..47B}, 
it seems likely that the magnetic field submerges together with the plasma in the outer penumbra. Na\"ively, we would expect to find magnetic fields of opposite polarity together with penumbral downflows.
As stated in \citet{2006A&A...447..343S}, 
however, even in high-resolution magnetograms by
\citet{2005A&A...436.1087L}, 
only a small fraction of the penumbral downflows show an opposite-polarity signal.

This observational lack of opposite polarities may be explained by the magnetic field returning below the surface in the very deep photosphere 
\citep{2001ApJ...547.1130W, 
2007PASJ...59S.593I, 
2011PhDT.......137F}. 
Since such a configuration creates asymmetric Stokes profiles
\citep[cf.][]{1975A&A....41..183I, 
1978A&A....64...67A, 
1992ApJ...398..359S}, 
the opposite-polarity signal is obscured even for disk center observation. Therefore it is important to use data of high spatial and spectral resolution to accurately capture line asymmetries and study penumbral return fields at least indirectly.

Recently, asymmetric Stokes V profiles with three lobes have been investigated, as they are a proxy for magnetic fields of opposite polarity in the lower penumbral photosphere 
\citep{2012AN....333.1009F, 
2012ApJ...750...62R, 
2013A&A...549L...4R, 
2013A&A...553A..63S, 
2013A&A...550A..97F}. 
These studies showed that a large portion of the penumbral downflows are accompanied by a magnetic field of opposite polarity. 
\citet{2013A&A...557A..25T,2015A&A...583A.119T} 
analyzed of the magnetic field configuration in sunspot penumbrae based on Stokes inversions and took into account the point spread function (PSF) that affects the data.

Observation in the near-infrared part of the solar spectrum provides another possibility of studying magnetic fields close to the solar surface. The reason is that the continuum absorption coefficient is lowest in the infrared (IR)
\citep{2002neko.book.....U}, 
and spectral lines at this wavelength are sensitive to the deepest parts of the solar photosphere
(\citealt{1992A&A...263..312S}; 
see~also \citealt{2014LRSP...11....2P} 
for a recent review on IR solar physics). Yet another benefit is that some of the neutral iron lines around 1.5~$\upmu$m have a high-excitation potential, causing them to be formed within a narrow photospheric layer. These lines are therefore less susceptible to gradients of the atmospheric parameters along the line of sight (LOS) and hence produce less asymmetric Stokes parameters \citep[see][for a comparison of the asymmetry of penumbral Stokes V profiles measured in the visible and the IR]{2002A&A...393..305M}. 
These characteristics make spectropolarimetry in IR lines an ideal tool for directly probeing penumbral magnetic fields of opposite polarity.

A drawback of IR observations is a decrease in spatial resolution when compared to data available in the visible regime, which means that larger telescopes are required for compensation. The \mbox{GREGOR} Infrared Spectrograph
\citep[GRIS,][]{2012AN....333..872C}, 
which has been in operation since 2014, is able to provide spectropolarimetric IR data of unprecedented quality that are suited for our study. In this contribution, we use 71 datasets from the visible and IR parts of the solar spectrum to study magnetic fields of opposite polarity in different layers of the penumbral atmosphere.


\section{Observations, data calibration and reduction}
\label{sec:obs}

For this study, we combined Hinode-SP data from the visible and GRIS data from the IR part of the solar spectrum.

\paragraph{\bf Data in the visible from Hinode-SP:}
The spectropolarimeter (SP)
\citep{2001ASPC..236...33L} 
of the solar optical telescope
\citep{2008SoPh..249..167T} 
onboard the Hinode satellite measures the full Stokes vector of Fe I 630.15 nm ($\rm{g}_{\rm eff}=1.67$) and Fe I 630.25 nm ($\rm{g}_{\rm eff}=2.5$). The slit of the SP covers $164\arcsec$ of the solar disk, while the scanning mechanism can displace the image by $\pm239\arcsec$
\citep{2013SoPh..283..579L}. 
Two-dimensional maps are obtained by scanning with the spectrograph slit across the target in steps of $\approx 0\farcs15$. Since the SP is designed to sample critically and since the pixel size along the slit corresponds to $\approx 0\farcs16$, the spatial resolution of the SP data is about $0\farcs32$. The spectral sampling of the SP is 2.15 pm and the spectral PSF has a full width at half maximum (FWHM) of about 2.5 pm, and has been determined prior to launch using a tunable laser
\citep{2013SoPh..283..579L}. 
For our study, we downloaded the raw data summarized in Table~\ref{Tab_A1} and reduced them, with the IDL routine sp\_prep.pro provided by the Solar Soft package \citep{2013SoPh..283..601L}.

\paragraph{\bf Data in the near-infrared from GRIS:}
With the integration of the scanning mechanism in April 2014, the IR spectropolarimeter of the new 1.5~m solar telescope GREGOR 
\citep{2001AN....322..353V, 
2012AN....333..796S, 
2012AN....333..810D} 
has become fully operational.
Using the GREGOR Adaptive Optics System
\citep[GAOS,][]{2012AN....333..863B}, 
we observed the disk passage of several active regions between April and July 2014  (see Table~\ref{Tab_A2} for a list of the sunspots observed in the IR). All datasets consist of raster scans with a length ranging from $12\farcs6$ to $50\farcs4$ (i.e., 100 to 400 scan steps) 
and a width of $59\farcs2$ (i.e., 470 pixel along the slit). They cover a 4~nm wide spectral region around 1565~nm. Polarimetric calibration was performed using the GREGOR polarimetric calibration unit
\citep{2012AN....333..854H}. 

Dark subtraction, flat fielding, and polarimetric calibration were applied to the raw data using the GRIS data pipeline that was derived from the TIP instrument 
\citep[see][for calibration steps included in the TIP pipeline]{1999ASPC..184....3C, 
1999ASPC..183..264M}. 
The raw images exhibit a fringe pattern that is most probably introduced by the pre-filter in front of the ferroelectric liquid crystals of the GRIS polarimeter or by the prisms that realign the beam to properly illuminate the IR camera  
\citep[cf.][for a theoretical study on fringes in spectropolarimetric instruments]{2003A&A...401....1S}. 
By subtracting a sinusoidal fit from the raw data, the fringes were significantly reduced a posteriori. Spikes in the spectrum were removed by applying a median filter to spectral regions where the intensity exceeded a certain threshold when compared to neighboring pixels.

\begin{figure}[htbp!]
	\centering
		\includegraphics[width=0.8\columnwidth]{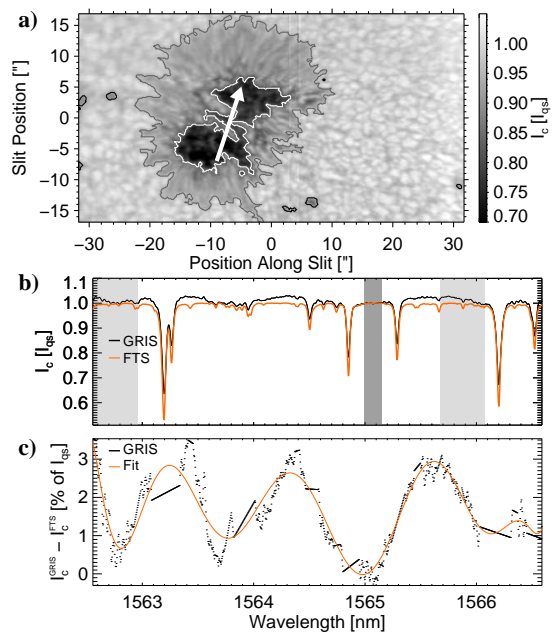}
		\caption{Panel a shows an intensity image of NOAA 12049 recorded with GRIS close to disk center. White and gray contours outline the umbra and penumbra. Black contours encircle regions that are not taken into account in the calculation of an average QS profile. The white arrow points to the center of the solar disk. In panel b we compare the average QS spectrum (black) with the FTS atlas (red). The regions shaded in gray are used in the analysis of the residual fringe pattern, see Sect.~\ref{sec:ana}. Panel c illustrates the intensity variation in the continuum of the GRIS spectrum (black dots) and a polynomial fit (red) used for correction.}
		\label{fig:Franz_fig00}
\end{figure}

\paragraph{\bf Comparison to FTS atlas:} As part of the calibration process, we created masks that define the quiet Sun (QS), penumbra, and umbra of the respective sunspot. To this end, we used the original data without any smoothing and applied certain criteria such as intensity and total polarization to define the respective boundaries. If automatic detection failed, for instance, because of intensity variations during data acquisition, we manually selected the umbral and penumbral areas. Finally, we defined all regions that contain strong magnetic fields, for example pores or orphan penumbrae, or regions that have been marked because there was no spectral information. This is because some maps from early observations exhibit vignetting caused by an aperture within the beam path of GREGOR that was placed there to minimize scattered light. When the image shift due to the correction of the tip-tilt mirror was large, the beam was partially blocked and the spectrograph slit was not entirely illuminated (cf. lower right corner of Fig.~\ref{fig:Franz_fig08} panel f). As an example we show the continuum intensity map of the following spot of active region NOAA 12049 on May 3 in Fig. \ref{fig:Franz_fig00} a. 

An average QS profile was computed from all the spectra outside of the above mentioned contours. We then compared this average QS spectrum of GRIS to an FTS atlas by
\citet{1991aass.book.....L} 
to obtain a correction curve for the continuum intensity in the Stokes I profiles, see Fig. \ref{fig:Franz_fig00} panels b and c. To exclude the spectral lines, we selected all measurements from the average QS spectrum at a wavelength where the intensity of the FTS atlas is $0.99 < \rm{I}_{\rm{FTS}} < 1.01$ (black dots). We then fit a polynomial (red curve) to the data points following an optimization procedure. The order of the polynomial was allowed to vary between 1 and 30, and we chose the polynomial with the smallest difference between measurement and fit.

\paragraph{\bf Spectral PSF:} Following
\citet{2004A&A...423.1109A} 
and
\citet{2007A&A...475.1067C}, 
we approximated the spectral PSF of GRIS.
To this end, we assumed that the PSF of the FTS is a delta function and degraded its spectrum until it overlapped with that of the average QS recorded by GRIS at disk center\footnote
{Since the FTS atlas was recorded at the center of the solar disk, this process only works well for disk center maps. With increasing heliocentric angle, this procedure becomes less applicable.}

\begin{equation}
\rm{I}^{\rm{GRIS}} = \rm{I}^{\rm{FTS}}_{\rm{convol}} = \frac{\rm{I}^{\rm{FTS}}+\EuScript{S}}{1+\EuScript{S}} \otimes \EuScript{G} (\rm{\omegaup})
\label{equ:convol}
\end{equation}

\noindent with $\rm{I}^{\rm{GRIS}}$ and $\rm{I}^{\rm{FTS}}$ being the QS spectra from GRIS and FTS respectively. $\EuScript{S}$ is the fraction of spectral scattered light and $\otimes$ represents a convolution. In first order, the spectral PSF is assumed to be a Gaussian $\EuScript{G}$ with FWHM of $\omegaup$. In a next step, we altered $\EuScript{S}$ and $\omegaup$ until the differences between $\rm{I}^{\rm{GRIS}}$ and $\rm{I}^{\rm{FTS}}_{\rm{convol}}$ were minimal.

\begin{figure}[htbp!]
	\centering
		\includegraphics[width=0.8\columnwidth]{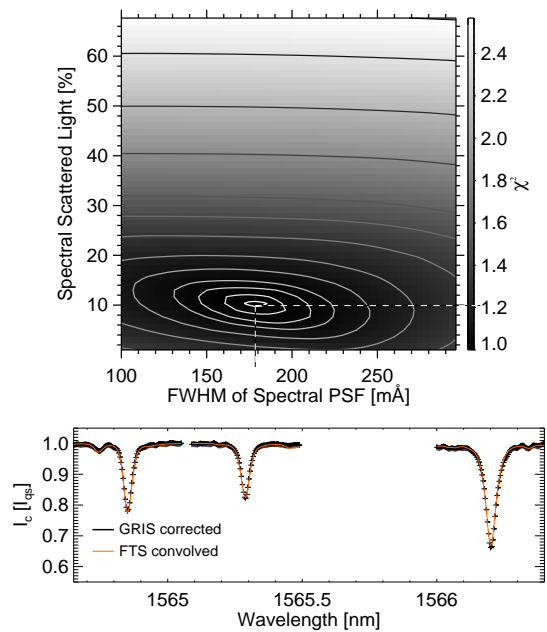}
		\caption{Top panel: Merit function that was used to determine the best values, i.e. FWHM and constant offset of a Gauss function, to model the PSF of GRIS.
Bottom panel: Three spectral lines that are used in this study (black crosses) overplotted with the same lines from the FTS atlas (red) after they were convolved with the spectral PSF of GRIS.}
		\label{fig:Franz_fig01}
\end{figure}

In the top panel of Fig. \ref{fig:Franz_fig01} we show the merit function for various $\EuScript{S}$ and $\omegaup$ for a particular dataset of May 3. A minimal $\upchi^2$, that is, the best agreement between GRIS measurement (black crosses) and convolved FTS spectrum (red line), is achieved for $\EuScript{S} = 10.3\%$ and $\omegaup = 17.8~\rm{pm}$. When we used these parameters for a Gauss function and convolved the FTS atlas with it, we obtaind the solid red line in the bottom panel of Fig. \ref{fig:Franz_fig01}. Except for a small discrepancy in the line core of Fe I 1566.20 nm, the agreement between the degraded FTS atlas and GRIS measurements is good.

\begin{figure}[htbp!]
	\centering
		\includegraphics[width=0.7\columnwidth]{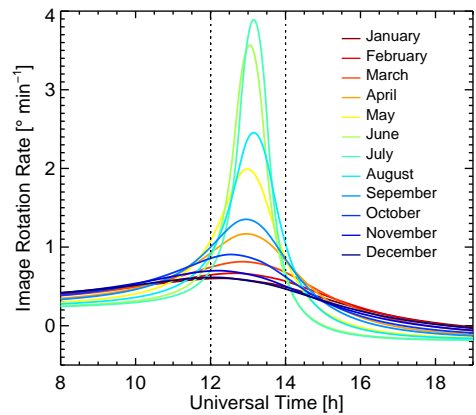}
		\caption{Rotation rate of the solar image in the focal plane of GREGOR. It shows the rotation rate on the first day of each month throughout 2014.}
		\label{fig:Franz_fig02}
\end{figure}

\paragraph{\bf Image rotation:} GREGOR is a telescope with an altitude-azimuth mount 
\citep{2012AN....333..816V},
which causes the solar image to rotate on the detector plane. The rotation rate across the spectrograph slit is time dependent. Figure ~\ref{fig:Franz_fig02} shows a model of the rotation throughout the day for the first day of each month. 

From April to September and between 12 UT and 14 UT, the image rotated more than $1^{\circ}$ per minute. No mechanical device was installed at GREGOR in 2014 to compensated for image rotation. This effect is visible in the image of panel a in Fig. \ref{fig:Franz_fig00} panel a, which was recorded at around 12 UT. The lock point of GAOS, which is kept at a fixed position, was on the light bridge, and the image rotates clockwise. Thus, toward the positive end of the slit, scanning and image rotation partly compensate for each other and the structures appear compressed perpendicular to the slit, see the elongated granules at $(x,y) = (30,15)$. Toward the negative end of the slit, scanning and image rotation have the same direction and the structures appear stretched, see the granules at $(x,y) = (-30,15)$. To avoid this distortion, we restricted our observations to times with an image rotation rate of less than $1^{\circ}$ per minute. This value is acceptable because a typical observation takes less than 20 minutes. 

One complication in the data evaluation process is due to the imperfect alignment of the optical and mechanical axes of the telescope. This causes the field of view to move along a complicated trajectory known as the beam wobble, when azimuth and elevation of the telescope are changed for the daily tracking of the Sun. During the observation this beam wobble is compensated for by GAOS, but it is very difficult to exactly translate the pointing coordinates of GREGOR into solar disk coordinates. To circumvent this problem we cross-correlated the scans with continuum images from the Helioseismic and Magnetic Imager 
\citep[HMI,][]{2012SoPh..275..229S} 
onboard the Solar Dynamics Observatory
\citep[SDO,][]{2012SoPh..275....3P} 
to determine the heliocentric angle of the sunspots observed with GRIS.


\section{Data description and analysis}
\label{sec:ana}

In our investigation, we combined information from the two Hinode Fe I lines in the visible with three IR Fe I lines from the GRIS spectra. Table~\ref{Tab_1} summarizes the selected lines alongside relevant parameters such as the wavelength $\lambdaup$ of the transition at ambient temperature and atmospheric pressure, atomic levels participating in the transition, excitation potential of the lower level $\chiup_{\rm{e}}$, and effective\footnote{The g-factor of only one level of the IR lines is known from laboratory measurements. The other is calculated assuming LS coupling.} Land\'e g-factor g$_{\rm{eff}}$. For a comparison of the sensitivity of the lines in Table~\ref{Tab_1} to various parameters such as temperature, velocity, and magnetic field at different atmospheric layers, we refer to \citet{2016A&A...Borrero}.

\begin{table}[h!]
\begin{center}
	\caption{Parameters of solar neutral iron lines. Information is taken from
\citet{1994ApJS...94..221N} 
except for g$_{\rm{eff}}$, which has been calculated by
$^{\rm{a}}$\citet{1986SoPh..107...57S} 
and $^{\rm{b}}$\citet{1995A&AS..113...71R}. 
}
	\begin{tabular}{cccc}
		\hline
		\hline
		\\[-2ex]
		{$\lambdaup $[nm]} & {multiplet $^{2\rm{S}+1}$L$_{\rm{J}}$} & {$\chiup_{\rm{e}}$ [eV]} & {g$_{\rm{eff}}$}\\	
		\\[-2ex]
		\hline		
		\\[-2ex]	
			{630.15} & {$^5\rm{P}^0_2 - \,^5\rm{D}_2$}& {3.65} & {1.67$^{\rm{a}}$} \\ 
			\\[-2ex]
			{630.25} & {$^5\rm{P}^0_1 - \,^5\rm{D}_0$}& {3.69} & {2.5$^{\rm{a}}$} \\ 
			\\[-2ex]							
			{1564.85} & {$^7\rm{D}_1 - \,^7\rm{D}^0_1$}& {5.43} & {3$^{\rm{b}}$} \\ 
			\\[-2ex]
			{1565.29} & {$^7\rm{D}_5 - [\frac{7}{2}]^0_4$} & {6.25} & {1.8$^{\rm{b}}$} \\
			\\[-2ex]
			{1566.20} & {$^5\rm{F}_5 - \,^5\rm{F}^0_4$} & {5.83} & {1.5$^{\rm{b}}$} \\ 
		\\[-2ex]
		\hline
		\hline
		\end{tabular}	
	\label{Tab_1}
\end{center}
\end{table}

\begin{figure}[htbp!]
	\centering
		\includegraphics[width=\columnwidth]{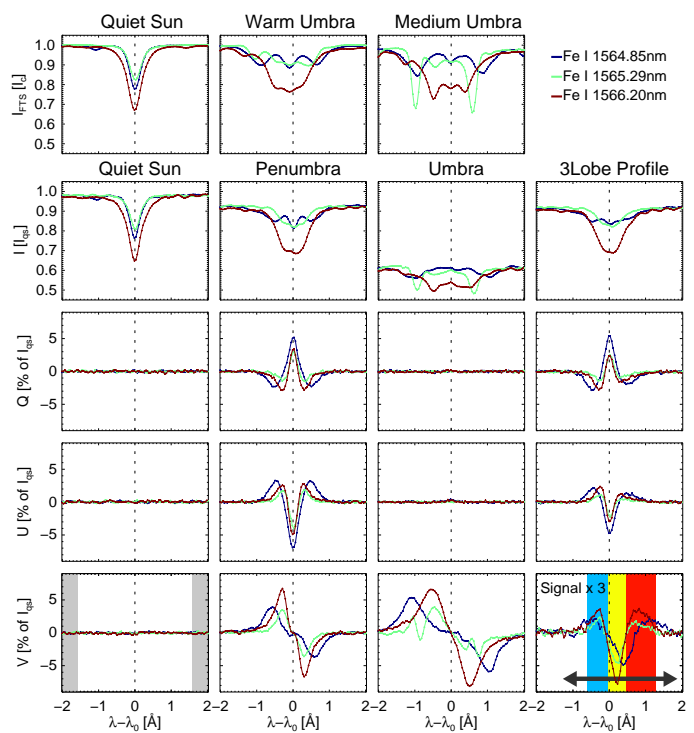}
		\caption{Stokes profiles of Fe 1564.85 nm (blue), Fe 1565.29 nm (green) and Fe 1566.20 nm (red) recorded with FTS and GRIS. Top row from left to right: Stokes I recorded with the FTS for the QS, a warm umbra and a medium warm umbra. Other rows from top to bottom: Stokes I normalized to the average QS, and Stokes Q, U and V in percent of the average continuum intensity. From left to right: Stokes vector from the QS, penumbra, umbra, and a region where the Stokes V profile shows three lobes. The Fe 1565.29 nm line is strongly blended by a hydroxyl radical line pair in the umbral example. The spectral regions used to estimate the noise in the polarimetric component are gray shaded in the lower left panel. The signal in the lower right panel is amplified by a factor of three and the spectral windows that are shifted across the spectrum to detect 3LPs are indicated in color. See main text for details.}
		\label{fig:Franz_fig03}
\end{figure}

\paragraph{\bf Example of measured IR Stokes profiles:} For our analyses, we extracted a 12~nm and a 40~nm wide spectral window around the Hinode-SP and GRIS lines respectively. As an example, we show Stokes profiles for the IR lines from different solar regions such as QS, penumbra, and umbra in Fig. \ref{fig:Franz_fig03}.

In the top row of Fig. \ref{fig:Franz_fig03} the intensity profiles from the FTS IR atlas are plotted
\citep{2001sus..book.....W} 
for the photospheric QS
\citep{1990Livingston}, 
a so-called warm umbra with an estimated field strength of 2000~G 
\citep{1983Giampapa}, 
and a so-called medium warm umbra with an estimated field strength of 2500~G 
\citep{1982Giampapa}. 
In the following, we discuss the features seen in the GRIS spectra, which are plotted in the other rows.  

For the QS, the continuum intensity of the GRIS Stokes I profile is slightly lower than that of the FTS atlas. The reason is that the former represents a pixel from an intergranular lane and is normalized to the intensity of the average QS, while the latter is an unresolved average QS profile by default because of the limited spatial resolution of the FTS. The resolving power of GRIS is high enough to identify weak lines such as the Fe~I blend at 1564.74~nm or Cr~I at 1566.34~nm, see red and blue profiles in the panels of Stokes I of Fig.~\ref{fig:Franz_fig03}. 

For penumbral Stokes I profiles, the $\upsigma$- and $\uppi$-components of the line triplet of Fe I 1564.85 nm are visible as a result of the high g-factor. The profiles Fe I 1552.29 nm and Fe I 1566.20 nm are significantly broadened. The polarimetric components show no peculiarities.

In the umbra, the Fe I 1564.85 nm line shows widely split $\upsigma$-components. The $\uppi$-component is absent, since the magnetic field is oriented along the LOS for an umbra observed close to disk center. The Fe I 1565.29 nm line profile is very shallow and displays several blends. The shallow profile is due to the high-excitation potential (Fe I 1565.29 nm shows the highest excitation potential of all IR lines considered in this study), which causes the lower level of the transition to be weakly populated at umbral temperatures. Additionally, the low temperatures allow for the formation of magnetically sensitive hydroxyl radicals (see green line in the third column of Fig. \ref{fig:Franz_fig03}). The $-\upsigma$ component is completely hidden by the blend of the hydroxy radical, while the $+\upsigma$ component remains visible at $-0.5 \AA$ from the line core. The influence of the hydroxyl radicals is also apparent in Stokes V, where additional lobes appear in both wings of Fe I 1565.29 nm. Because of this, information from this line has to be taken with caution for spectra from the inner penumbra and umbra. The Fe I 1566.20 nm line profile is less split and significantly deeper than the other two lines. In the umbral spectra, there are CN and CO blends apparent around 1566.06~nm (red line in the third column of Fig. \ref{fig:Franz_fig03}).

In the fourth column, we plot a profile from the penumbra where Stokes V shows three lobes (the signal is amplified by a factor of three). The shape is similar in all three lines, only the broadening is higher in Fe I 1564.85 nm that in the other lines.

\paragraph{\bf Residual fringes:} One feature still visible in the intensity profile of the GRIS data are the wiggles in the continuum, which originate from the fringes attenuated during the reduction process. Modeling the fringes by a superposition of sine curves of various amplitudes and frequencies requires little computing time when compared to more involved approaches using wavelet analysis
\citep{2006ApJ...649..553R} 
or 2D pattern recognition
\citep{2012ApJ...757...45C}. 
Drawbacks are the danger of removing or modifying an actual signal, that is a spectral line. Therefore, the filter was applied in a conservative way, accepting residual fringes in the data. 

\begin{figure}[htbp]
	\centering
		\includegraphics[width=\columnwidth]{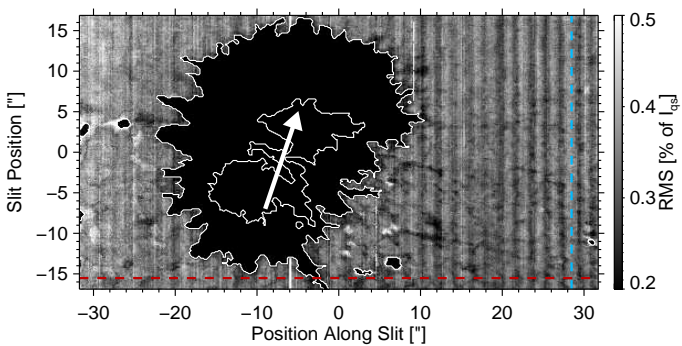}
		\caption{Spatial distribution of residual fringes. The figure shows the RMS fluctuations around the continuum intensity at $\uplambda = 1565.07~\rm{nm} \pm 0.8~\rm{nm}$ in GRIS data. Contours outline umbra, penumbra, and some nearby concentrations of magnetic field. The white arrow points to the center of the solar disk. The horizontal (dashed red) and vertical (dashed blue) lines mark regions where we investigated the residual fringes along the slit (Fig.~\ref{fig:Franz_fig05}) and for different slit positions (not shown).}
		\label{fig:Franz_fig04}
\end{figure}

To study the residual fringes in the spatial dimension, we subtracted the degraded FTS profile from each Stokes I profile (see Eq.~\ref{equ:convol}). Then we computed the RMS fluctuation in a region without lines. We chose $\uplambda = 1565.07~\rm{nm} \pm 0.08~\rm{nm}$ as indicated by the dark gray region in Fig.~\ref{fig:Franz_fig00} panel b. Figure~\ref{fig:Franz_fig04} shows the spatial distribution of the fluctuation. The fluctuations were manually set to zero at the location of the sunspot. The columns at $-6\arcsec$ and $+9\arcsec$ are saturated because at this position along the slit a spike increases the RMS value significantly. The resulting pattern of the RMS values is stable for different scan positions, but varies along the slit with an amplitude of $(3.5\pm1.5)\cdot10^{-3}\cdot\rm{I}_{\rm{qs}}$.

To visualize the spectral dependence of the fringes for all pixels along the slit, we removed the intensity variation that is due to granulation. This was done by subtracting the mean Stokes~I value in the interval $\uplambda = 1562.76~\rm{nm} \pm 0.2~\rm{nm}$ and $\uplambda = 1565.48~\rm{nm} \pm 0.2~\rm{nm}$, as indicated by the light gray regions in Fig.~\ref{fig:Franz_fig00} panel b. 

\begin{figure}[htbp]
	\centering
		\includegraphics[width=\columnwidth]{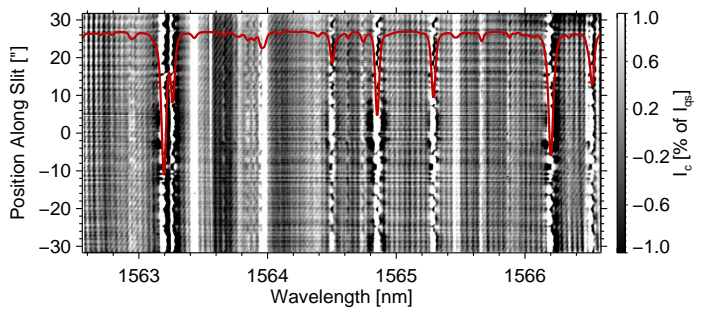}
		\caption{Spectral distribution of residual fringes. The figure shows the difference between GRIS Stokes I spectra and the degraded FTS profile (red line) along the slit. The GRIS Stokes I spectra have been corrected for granular intensity variation.}
		\label{fig:Franz_fig05}
\end{figure}

The fringes have a peak-to-peak amplitude of approximately 1\% and are highest at the lower and upper wavelength end of the spectrum. In addition to the vertical fringe pattern, a second and weaker diagonal fringe pattern is visible along the slit (see Fig.~\ref{fig:Franz_fig05}). The wavelength position of the fringes was stable during the scan, but the amplitude changed, possibly as a result of the varying seeing conditions. 

\paragraph{\bf Polarimetric accuracy:} To ensure that no residual fringes remain in the polarimetric parameters that might compromise our analysis, we estimated the noise in Stokes Q, U and V. To this end, we calculated maps of the maximum of the total polarization (P$_{\rm{tot}}$) of the lines listed in Table~\ref{Tab_1}, 
\begin{equation}
\rm{max}~\rm{P}_{\rm{tot}} = \rm{max} \left[ \sqrt{\rm{Q(\uplambda)}^{2}+\rm{U(\uplambda)}^{2}+\rm{V(\uplambda)}^{2}} \right]_{\uplambda_{\rm{blue}}}^{\uplambda_{\rm{red}}}
\end{equation}
in which Q, U, and V denote the polarimetric Stokes parameters normalized to the continuum intensity of the average QS, while $\uplambda_{\rm{blue}}$ and $\uplambda_{\rm{red}}$ represent the lower and upper limits of the spectral window under study. In a subsequent step, we extracted all Stokes V spectra that belong to the QS, that is, profiles for which max~$\rm{P}_{\rm{tot}} \le 5 \cdot 10^{-3} \cdot \rm{I}_{\rm{c}}$. If the number of profiles in the resulting subset was not sufficient for averaging (we chose a minimum of 10000 profiles), the threshold of max~$\rm{P}_{\rm{tot}}$ was increased to $7 \cdot 10^{-3} \cdot \rm{I}_{\rm{c}}$ or $1 \cdot 10^{-2} \cdot \rm{I}_{\rm{c}}$, respectively. In a last step, we calculated the RMS fluctuation of the Stokes V spectra of the subset in the continuum regions. In Hinode-SP we defined the continuum as all the measurements that are at least $\pm 37$~pm detuned from the line core. For the GRIS data we chose a minimal detuning of $\pm$~160~pm shown as gray shaded areas in the lower left panel of Fig.~\ref{fig:Franz_fig03}.

The average of the RMS fluctuation in the continuum region of all QS Stokes V spectra was then, in turn, interpreted as the 1$\upsigma$ noise level. It varies from $8.3$ to $15.3\cdot10^{-4}\cdot\rm{I}_{\rm{c}}$ (see Table \ref{Tab_A1}) in the Hinode data and ranges\footnote{This is due to various combinations of exposure time (30~ms to 100~ms) and number of accumulations (3 to 10).} from $6.8$ to $11.2\cdot10^{-4}\cdot\rm{I}_{\rm{c}}$ (see Table \ref{Tab_A2}) in the GRIS data. Similar analyses were performed for Stokes Q and U. We did not find any indication of the fringe pattern in the polarimetric components above the 3$\upsigma$ noise level.


\begin{figure*}[htbp!]
	\centering
		\includegraphics[width=\textwidth]{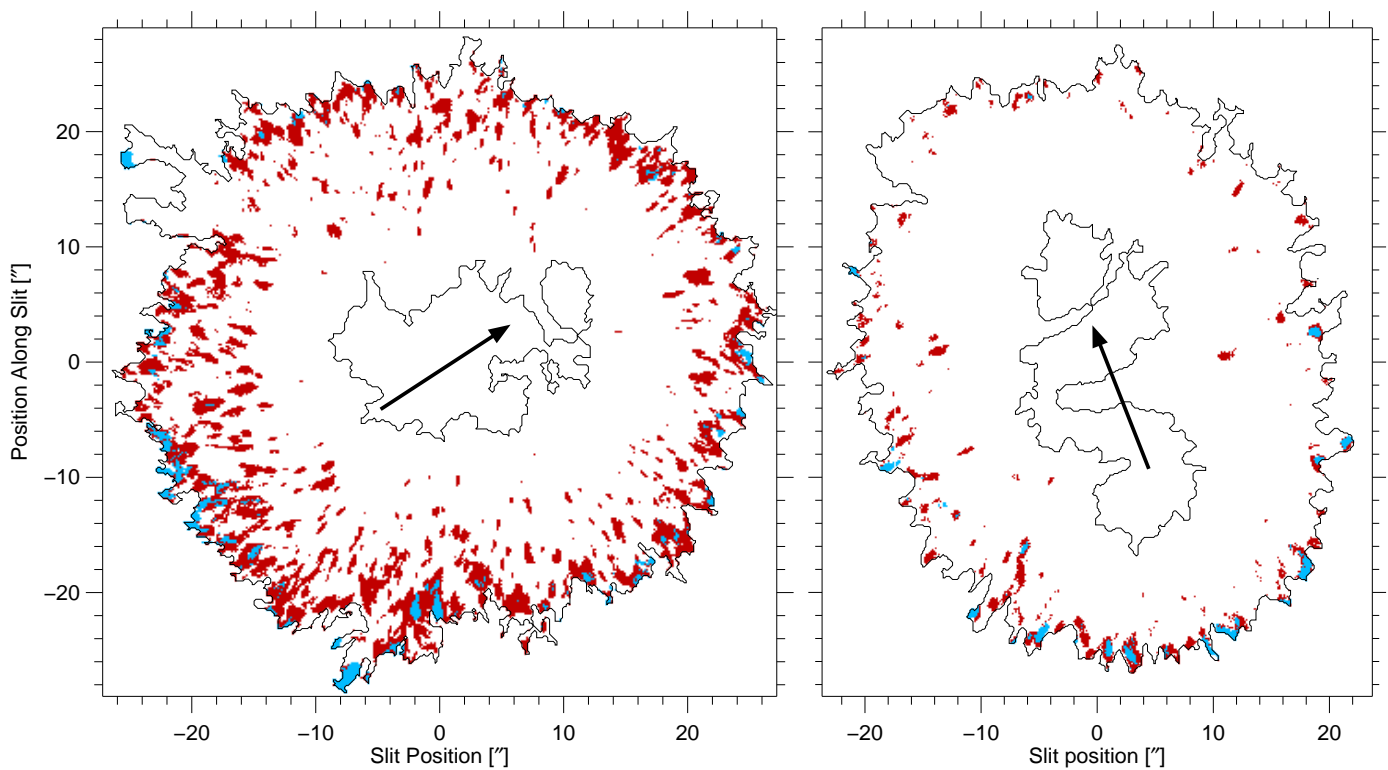}
		\caption{Distribution of three-lobed (red) and reversed-polarity (blue) Stokes V profiles in the penumbra. The left panel shows sunspot NOAA 10933 recorded on January 5, 2007 with Hinode-SP at $\Uptheta\approx3^{\circ}$. The right panel shows NOAA 12049 recorded on May 3, 2014 with GRIS at $\Uptheta\approx6^{\circ}$ (right). The contours outline the penumbral boundaries and the black arrows point to the center of the solar disk.}
		\label{fig:Franz_fig06}
\end{figure*}

\section{Results and discussion}
\label{sec:res}

We define a 3LP as a regular Stokes V profile with a third lobe on its red wing that is of the same polarity as the blue lobe of the regular Stokes V profile
\citep[cf.][for another type of three-lobed profile]{2005A&A...436..333B,
2010A&A...524A..20K}. 

To detect such profiles, we applied the procedure described in
\citet{2013A&A...550A..97F}, 
which are summarized in the following. First, we corrected for the polarity of the sunspot in such a way that the blue lobe of the V profile was always positive. To analyze the Hinode-SP data, we extracted both Fe lines from the spectrum to be able to treat them separately. Then, we split the Stokes V profile into three spectral windows. The first two windows have a bandwidth of 20~pm, while the third is 45~pm wide. The position of these spectral windows may vary by 35~pm, but neither their width nor their order was allowed to change. This is indicated by the colored regions in the upper right panel of Fig.~1 in 
\citet{2013A&A...550A..97F}. 
Because the IR lines are broader, we increased the width of the spectral windows, allowing us to employ the same algorithm as we used to work with GRIS data. The modified windows are indicated by the colored regions in the lower right panel of Fig.~\ref{fig:Franz_fig03}. They have a width of 60~pm (blue), 40~pm (yellow), and 80~pm (red). As before, their position may vary by $\pm$60~pm (indicated by the black arrow), but neither their width nor their order was allowed to change. 

Finally, a Stokes V profile is counted as a 3LP when the following conditions apply to the extrema of the positive and negative lobes and to their adjacent spectral measurements: 
\begin{itemize}
\item [a)] {They are positive in the blue and red windows, but negative in the yellow window.}
\item [b)] {They are above the 3$\cdot\upsigma$ noise level in all windows.}
\end{itemize}

As an example of our results, Fig.~\ref{fig:Franz_fig06} compares a Hinode-SP dataset (Fe~I~630.25~nm) to a dataset recorded with GRIS (Fe~I~1566.2~nm). Even though the data were not recorded simultaneously, a line-up like this is justified because the data are very similar in terms of sunspot size (1900~arcsec$^2$ vs. 1450~arcsec$^2$), heliocentric angle ($\Uptheta\approx3^{\circ}$ vs $\Uptheta\approx6^{\circ}$), spatial resolution (0\farcs32 vs. $\approx$ 0\farcs4), spectral sampling (2.15~pm vs. 4.0~pm), and polarimetric accuracy (9.8$\cdot10^{-4}$ in both cases). The main differences are the observatory itself (space-borne vs. ground-based) and spectroscopic line (visible vs. IR). The latter implies a different sampling of the solar atmosphere. The Hinode-SP lines in the visible are sensitive to a broad atmospheric layer reaching up to the mid-photosphere.The formation of the IR lines of GRIS, on the other hand, occurs in a narrow, very deep layer of the photosphere, see Fig. 9 in 
\citet{2016A&A...Borrero}. 
However, this is a simplification since the two lines have different temperatures and different Zeeman sensitivities.

We indicate the regions with 3LPs in red and regions with reversed-polarity Stokes V profiles (RPPs) in blue. In both datasets, 3LPs appear predominantly in the outer penumbra, with some patches also present in the middle penumbra. For Fe~I~630.25~nm, however, they are more frequent and larger than for Fe~I~1566.2~nm. The largest patch in the left panel of Fig.~\ref{fig:Franz_fig06} at (x,y)=(-2,-22) has a diameter of 5\arcsec, while it is only 2\arcsec~for the largest patch in the right panel at (x,y)=(3,-25). An excess of 3LPs is visible on the limb-side penumbra of both datasets. Several narrow and elongated regions of 3LPs with a width of about 0\farcs5, for instance, at (x,y)=(18,12), are visible throughout the penumbra in the Fe~I~630.25~nm data. Structures like this are very rare in Fe~I~1566.2~nm and completely absent from the center-side penumbra. Regions of RPPs are only located in the outer penumbra and are similar in terms of morphology and abundance for both spectral lines. The majority appears close to regions of 3LPs, some of them are even entirely encircled by them, for instance, \mbox{(x,y)=(12,-19)} in the left panel and \mbox{(x,y)=(3,19)} in the right panel, while others are completely independent of  3LPs, for example,  \mbox{(x,y)=(-8,-25)} in the left panel and \mbox{(x,y)=(18,-9)} in the right panel.

\begin{table}[h!]
\begin{center}
	\caption{Comparison of penumbral area showing RPPs and 3LPs together with the amount of total area showing opposite polarity as measured in visible and IR lines.}
	\begin{tabular}{cccc}
		\hline
		\hline
		\\[-2ex]
		{line} & {reversed polarity} & {three-lobed} & {tota}l\\	
		{$\uplambda$ [nm]} & {area [\%]} & {area [\%]} & {area [\%]}\\	
		\\[-2ex]
		\hline		
		\\[-2ex]	
			{630.15} & {2.4} & {11.4}  & {14.0}\\ 
			\\[-2ex]
			{630.25} & {1.9} & {15.1}  & {17.0}\\ 
			\\[-2ex]							
			{1564.85} & {1.1} & {0.7}  & {1.8}\\ 
			\\[-2ex]							
			{1565.29} & {1.1} & {0.6}  & {1.7}\\ 
			\\[-2ex]							
			{1566.20} & {0.9} & {3.2}  & {4.1}\\ 
		\\[-2ex]
		\hline
		\hline
		\end{tabular}	
	\label{Tab_2}
\end{center}
\end{table}

For the other lines under study, we find similar results, which we summarize in Table~\ref{Tab_2}. In the visible, a few percent of the penumbral area are covered by RPPs, while 3LPs occupy an additional 12\% to 15\%, which means that up to 17\% of the penumbral area is covered by magnetic fields of opposite polarity. In the IR the numbers for the RPPs are about a factor of two smaller, while the amount of 3LPs is less than a few percent and thus up to an order of magnitude smaller than in the visible.

The lack of penumbral 3LPs in the IR is expected from models explaining such profiles as due to a magnetic field of reversed polarity in the lower part of a two-layer atmosphere \citep[see Fig.~\ref{fig:Franz_fig08} panel a and Fig.~7 in][]{2013A&A...550A..97F}. 
Within such a scenario, the visible line would be sensitive to both atmospheric layers, but the IR line only to the lower one. As a result, IR data should show fewer 3LPs, but since the total amount of magnetic fields of reversed polarity should stay constant, we would expect an increase in RPPs. However, as illustrated by the blue area in Fig.~\ref{fig:Franz_fig06} and summarized by the numbers in Table~\ref{Tab_2}, we did not find such an increase in any line of the IR data. As a consequence, the amount of penumbral area covered by magnetic fields of opposite polarity in Fe~I~1566.2~nm appears up to one order of magnitude lower than Fe~I~630.25~nm. The disagreement of the numbers summarized in Table~\ref{Tab_2} might be due to several reasons. We mention spatial stray light, problems with the detection algorithm, and selection bias or poor statistics. Spatial stray light, resulting from a combination of atmospheric seeing and instrumental effects, smears out signals over a certain region. This dilutes the signature of a localized 3LP
\citep[cf., e.g.,][]{2016A&A...Lagg}, 
making a detection more difficult. An analysis of the influence of spatial stray light is quite involved, especially if the time-dependent PSF is unknown, and will therefore be treated in a separate publication. In addition the higher Zeeman sensitivity in the infrared, which requires larger velocity gradients to produce asymmetric Stokes V profiles at 1.5 $\mu$m than in the visible
\citep{1989A&A...221..338G}, 
might influence on the amount of detected 3LPs. In the following, we investigate the influence of the some of the above-mentioned phenomena on the detection of 3LPs.

\paragraph{\bf Comparison of 3LPs in visible and IR lines:}First, we address the question whether the strategy we adopted to detect 3LPs in visible data is applicable to IR data. We argue that 3LPs in the visible and in the IR look very similar and thereby justify the modifications to the algorithm described above. Since there are no simultaneous observations of Hinode and GREGOR for a sunspot at disk center, we chose a theoretical approach of inversion and subsequent forward modeling. We revisited the downflow case described in
\citet{2013A&A...550A..97F}, 
where we inverted a 3LP of the Fe I 630.25 nm line with SIR 
\citep{1992ApJ...398..375R}. 
Then we used the resulting atmosphere to synthesize a 3LP as would be observed by GRIS for Fe~I~1566.2~nm.

\begin{figure}[h!]
	\centering
		\includegraphics[width=0.8\columnwidth]{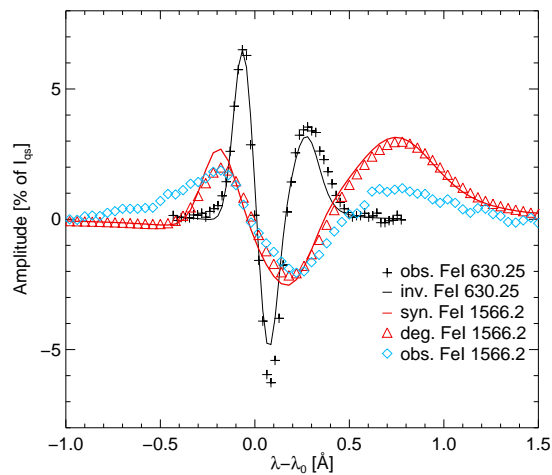}
		\caption{Shape of 3LP in visible and IR lines. A profile measured by Hinode-SP (black crosses) is drawn alongside the results of a SIR inversion (solid black). The corresponding model atmosphere is used to synthesize a profile in the IR at 1566.2 nm (solid red), which is shown together with the subsequently degraded profile as would be measured by GRIS (red triangles). For comparison, we show a 3LP from a GRIS observation (blue diamonds).}
		\label{fig:Franz_fig07}
\end{figure}

A summary of our results is shown in Fig.~\ref{fig:Franz_fig07}, where we plot a 3LP from Hinode-SP measurements of Fe I 630.25 nm (black crosses) alongside the output profile from a SIR inversion (solid black) of that line. The inversion reproduces all the features of the observed profile, although the amplitude of the negative lobe is smaller in the inversion than in the measurements. A large number of nodes might lead to better agreement, but our aim was to reproduce all features with the fewest number of free parameters.

In a next step, the model atmosphere that corresponds to this output profile was used to synthesize the Stokes V profile of the Fe~I~1566.20~nm IR line (solid red). Both Stokes V profiles exhibit three lobes and, at least for this example, this demonstrates that an atmosphere that causes a 3LP for Fe~I~630.25~nm yields a 3LP in Fe~I~1566.20~nm. In the following, we discuss the details of these 3LPs.

When compared to the Stokes V profile in the visible, the resulting IR profile is significantly broader. This is mainly caused by the Zeeman effect, which has a quadratic wavelength dependence, and the Doppler broadening, which depends linearly on wavelength. For Fe~I~630.25~nm, the profile shows a positive and a negative lobe with similar amplitudes located symmetrically around the zero crossing wavelength $\uplambda_{0}$. A third lobe with positive polarity but smaller amplitude is located 0.3$\AA$ on the red side of $\uplambda_{0}$. For the synthetic Fe~I~1566.2~nm, there are a positive and a negative lobe with the same amplitudes located around $\uplambda_{0}$. The negative lobe, however, is slightly wider than the positive lobe. The third lobe is located at $+0.75$$\AA$ on the red side of $\uplambda_{0}$ and shows a larger amplitude than the other two lobes.

The difference between the inverted and the synthesized 3LPs may be explained by the configuration of the model atmosphere, that is, by an increase of inclination and strength of the magnetic field with optical depth. The two lobes around $\uplambda_{0}$ are more sensitive to the higher atmospheric layers, while the third lobe on the red side of the profile receives more contribution from the lower atmospheric layers. Since the Fe~I~1566.2~nm is more sensitive to these lower layers, a larger amplitude for the third lobe in Fe~I~630.25~nm can be expected. This can can only be considered an estimate because a model atmosphere resulting from the inversion of one line always falls short of representing the other, for instance, because they do not sample exactly the same range of heights and have different sensitivities to temperature, magnetic field, etc.

In a final step, we degraded the synthetic IR profile with the spectral PSF calculated in Sect.~\ref{sec:obs}, thereby simulating a profile as would be recorded by GRIS (red triangles). Except that the amplitude of the peaks is slightly diminished, there is no significant change. When we plot the simulated IR profile alongside a profile as observed by GRIS (blue diamonds), we find that the third lobe is largely overestimated by the synthetic profile, but the width and wavelength position of the individual lobes are reproduced correctly. We performed this test with Fe~I~1564.8~nm and Fe~I~1565.2~nm and obtained similar results. This comparison has to be taken with caution as the simulated and observed GRIS profiles represent the atmospheric conditions in different spots observed at different times. Nevertheless, it shows that the IR lines do indeed exhibit 3LPs and that the modification of the spectral windows in our detection algorithm is reasonable and does not cause of the lack of 3LPs in the IR.

\paragraph{\bf Geometrical considerations:} 
\citet{2013A&A...550A..97F}, 
used the synthesis module of SIR to study the effect of atmospheric parameters on the shape of Stokes V profiles. Our goal was to create a simple model that yields 3LPs with a non-vanishing area asymmetry. For all the realizations where we succeeded, the overall configuration of the model was the following:

\begin{itemize}
\item [1)] {An atmosphere consisting of two layers.}
\item [2)] {Magnetic fields in both layers, but of opposite polarity.}
\item [3)] {Plasma moving away from the observer in one layer.}
\end{itemize}

Our observations show that Stokes V profiles with three lobes are accompanied by asymmetric Stokes I profiles in which the line wing is redshifted, while the line core appears unshifted. Owing to the temperature stratification of the solar photosphere, information about the lower atmospheric region is encoded in the wings of the spectral lines, while the core of the line is in turn more sensitive to higher atmospheric regions
\citep{1978stat.book.....M, 
2003isp..book.....D}. 
Furthermore, plasma motions occur predominantly in the same direction as the magnetic field
\citep{2003A&A...403L..47B}. 
Thus, it is reasonable to assume that both the plasma flow and the magnetic field of opposite polarity are present in the lower atmospheric layer. 

\begin{figure}[htbp!]
	\centering
		\includegraphics[width=\columnwidth]{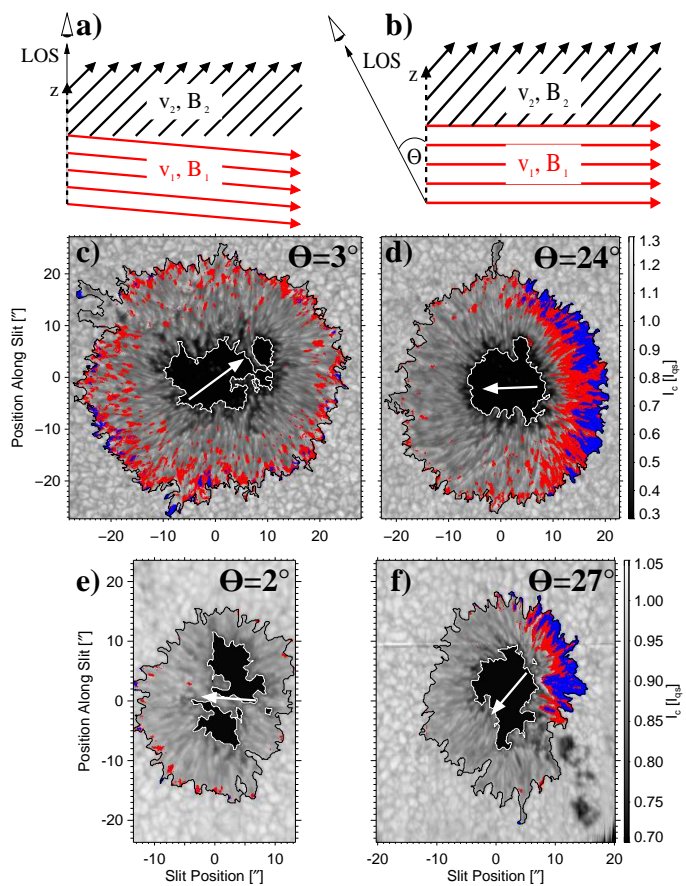}
		\caption{Top: Sketch of a simple atmospheric model capable of producing 3LPs in penumbrae at disk center (panel a) and larger heliocentric angles (panel b). B$_1$ , B$_2$, v$_1$ and v$_2$ represent magnetic field strength and velocity in the lower and upper atmospheric layer. The LOS is parallel to the local vertical z at disk center and seen under the projection of $\Uptheta$ for increasing heliocentric angles. Middle: Hinode-SP intensity image of a sunspot at disk center (panel c) and off center (panel d). Regions with 3LPs are indicated in red, and regions with RPPs are shown in blue. Bottom: Same as above, but for GRIS observations. The black and white contours indicate the penumbral boundaries, and the white arrow points to the center of the solar disk.}
		\label{fig:Franz_fig08}
\end{figure}

Figure~\ref{fig:Franz_fig08} shows an atmospheric configuration similar to that of
\citet{1993A&A...275..283S}, 
which was modified in such a way that it yields 3LPs in penumbrae at disk center, panel a, and at larger heliocentric angles, panel b. We drew the magnetic field and the velocity of the plasma in the lower atmospheric layer, moving away from the observer, in red. 

At disk center, the combination of penumbral downflows and magnetic fields of opposite polarity causes 3LPs in the outer penumbra (see panels c and e of Fig.~\ref{fig:Franz_fig08}). At larger heliocentric angles, projection effects have to be considered. For sunspots away from disk center, the horizontal Evershed outflow, which is present in the lower layers
\citep{1964ApNr....8..205M} 
of the penumbral intraspines
\citep{1993ApJ...418..928L, 
2007PASJ...59S.593I}, 
causes a red-shift of the spectral line in the limb-side penumbra. While the more vertical magnetic field in the penumbral spines
\citep{1993ApJ...418..928L} 
still shows the same polarity as the umbra, the horizontal fields of the intraspines
\citep{2004A&A...427..319B} 
show reversed polarity at and beyond the magnetic neutral line. The combination of these two effects yields 3LPs and explains why they appear predominantly in the dark filaments of the limb-side penumbra 
\citep[see Fig.~\ref{fig:Franz_fig08} panels d and f and cf. also][]{1992ApJ...398..359S, 
2009A&A...508..963S}. 

\paragraph{\bf Center-to-limb variation of 3LPs:} So far, we have described only two different sunspots at disk center that were recorded individually by Hinode-SP and GRIS. To rule out that the resulting discrepancy of 3LPs in the visible and in the IR is due to selection bias, we performed a statistical analysis. We included 71 sunspot maps from the Hinode-SP and GRIS data archives, which were recorded at various heliocentric angles. From these data we computed the center-to-limb variation (CLV) of the penumbral area covered by 3LPs and RPPs. This allowed us to extrapolate the amount of 3LPs and RPPs to $\Uptheta = 0^{\circ}$, where the opposite polarity signal is exclusively due to the penumbral magnetic field returning into the Sun. The drawback of this procedure is that away from disk center, 3LPs and RPPs may be caused by projection effects and are not necessarily proxies of opposite polarities anymore. This has to be kept in mind when studying the CLV of the area coverage of penumbral 3LPs. 

\begin{figure}[htbp!]
	\centering
		\includegraphics[width=\columnwidth]{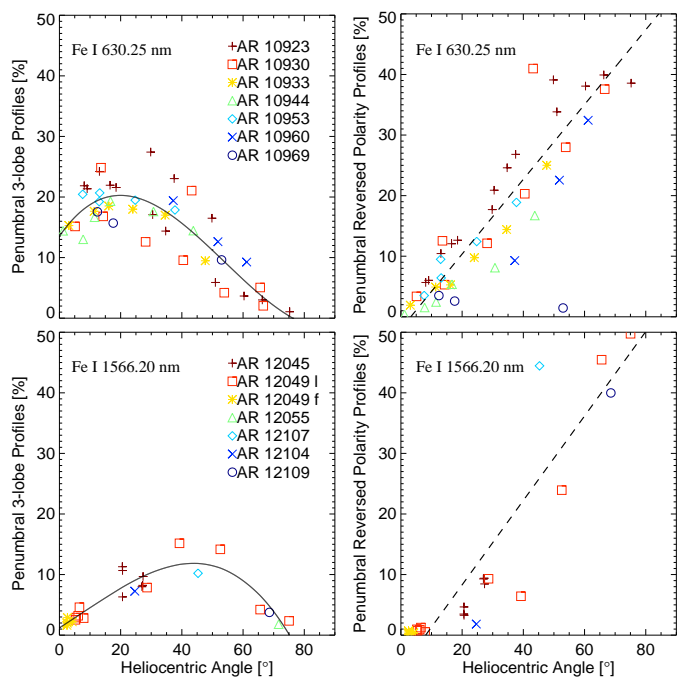}
		\caption{Center-to-limb variation of the amount of penumbral 3LPs (left) and RPPs (right). Results for Fe~I~630.25~nm and Fe~I~1566.2~nm are depicted in the top and bottom panels respectively. Each symbol represents a particular sunspot, followed during its disk passage. To extrapolate the values to disk center, we fit a third-order polynomial to the data points of 3LPs (solid black) and a straight line to the RPPs (dashed black).}
		\label{fig:Franz_fig09}
\end{figure}

The top left panel of Fig.~\ref{fig:Franz_fig09} shows the results for the Fe~I~630.25~nm line observed by Hinode-SP. Each symbol represents the area coverage of 3LPs normalized to the area of the penumbra for one sunspot at various heliocentric positions. Different symbols represent the disk passage of different sunspots. For example, for AR 10923 (red crosses) we found 21\% 3LPs at $\Uptheta=9^{\circ}$, reaching a maximum of approximately 25\% at $\Uptheta\approx20^{\circ}$ before their amount gradually decreases to 1\% at $\Uptheta=75^{\circ}$. For all the spots in our survey, we found a similar behavior: from disk center to $\Uptheta \approx 25^{\circ}$, we measured an increase of the amount of penumbral 3LPs before numbers decrease and vanish around $\Uptheta=75^{\circ}$. To illustrate this relationship, we fit a third-order polynomial (solid black line in the top left panel of Fig.~\ref{fig:Franz_fig09}) to the measurements. The scatter of the individual values is fairly large, for instance, $\pm5\%$ at $\Uptheta\approx30^{\circ}$, which hampers correctly deduceing the dependence of the amount of 3LPs on $\Uptheta$.

In the top right panel of Fig.~\ref{fig:Franz_fig09} we plot the fractional area coverage of RPPs normalized to the penumbral area. For AR 10923 (red crosses), we found that the amount of RPPs depends almost linearly on the heliocentric angle. For $\Uptheta=9^{\circ}$ we found 6\%, which gradually increases to 40\% at $\Uptheta=75^{\circ}$. The other sunspots showed a similar trend, which we visualized assuming a linear behavior (dashed black line in the top left panel of Fig.~\ref{fig:Franz_fig09}). Again, the individual scatter is significant, for instance, $\pm10\%$ for AR 10930 at $\Uptheta\approx40^{\circ}$. This explains why the small numbers of RPPs for $\Uptheta < 5^{\circ}$ are not correctly modeled by the fit.

In the bottom left panel of Fig.~\ref{fig:Franz_fig09}, we plot the results of the Fe~I~1566.2~nm line observed by GRIS. Again, each symbol represents the amount of penumbral 3LPs during its disk passage, while different symbols represent the different sunspots. The GRIS dataset covers fewer sunspots than the Hinode dataset, but contains more observations at disk center. We distinguish between a leading (AR 12049 l) and a following (AR 12049 f) spot. For the leading spot (orange squares), we found 3\% 3LPs at $\Uptheta=5^{\circ}$, a maximum of 15\% at $\Uptheta\approx40^{\circ}$, and 2\% at $\Uptheta=75^{\circ}$. As before, we fit a third-order polynomial (solid black) to the dataset to visualize the trend in the IR data. This curve looks significantly different from the Fe I 630.25 nm data. It increases gradually, peaks at $\Uptheta=45^{\circ}$ before it drops more rapidly, vanishing at $\Uptheta=75^{\circ}$. The maximum of the penumbral area occupied by 3LPs is 12\% instead of 20\% as measured in the Fe~I~630.25~nm line at $\Uptheta \approx 20^{\circ}$. The scatter here, i.e. $\pm3\%$ at $\Uptheta\approx20^{\circ}$, is slightly smaller than in the Hinode data.

In the bottom right panel of Fig.~\ref{fig:Franz_fig09} we plot the penumbral area covered by RPPs in the GRIS data. For AR 12049 l (red crosses) we found 2\% at $\Uptheta=6^{\circ}$ which increases to 50\% at $\Uptheta=75^{\circ}$. The other spots show a similar trend, which we \mbox{visualized} using a linear fit (dashed black line). Again, the small numbers of RPPs at disk center are not modeled correctly for $\Uptheta < 5^{\circ}$. The slope of the linear fit is slightly steeper than for the Hinode data, but it is important to note that there is no qualitative difference between the CLV curves of the amount of RPPs in Fe~I~630.25~nm and Fe~I~1566.2~nm.

The reasons for the variance of the data points in Fig.~\ref{fig:Franz_fig09} are manifold. If the penumbra is asymmetric or only part of the spot was scanned (see Figs.~\ref{fig:Franz_figA01} and \ref{fig:Franz_figA02} in the Appendix), one side may appear larger than the other. This means that the fractional area covered by 3LPs is higher or lower, depending on whether the limb- or the center-side penumbra appears larger in the dataset. Furthermore, the Wilson depression results in a foreshortening of the center-side penumbra, causing a systematic increase in the amount of 3LPs and RPP. Sometimes the algorithm does not detect the boundaries of the penumbra accurately (see Fig.~\ref{fig:Franz_figA01} in the Appendix), which diminishes the correct value of 3LPs. Finally there are magneto-optical effects, which manifest themselves as two additional lobes of opposite polarity around the zero crossing position of Stokes V profiles in Fe~I~630.25~nm (see Fig.~\ref{fig:Franz_figA02} in the Appendix). This leads to an incorrect detection of 3LPs in the inner penumbra of large sunspots with a strong magnetic field. A manual correction of these effects is not necessary because they do not change the overall picture and would be subjective.
 

\section{Summary, conclusion and outlook}
\label{sec:con}

We used observations of the infrared spectropolarimeter of GREGOR and of the spectropolarimeter onboard Hinode to investigate penumbral magnetic fields of opposite polarity at different atmospheric heights. For the infrared data, we determined the spectral point spread function (PSF) and estimated the influence of image rotation on the data quality. Additionally, we studied the spatial and spectral distribution of residual artifacts, that is, spectral fringes, which are not completely removed during the calibration process. We found that they have a peak-to-peak amplitude of approximately 1\% in Stokes I but well below $10^{-3}$ in the polarimetric components. We concluded that these residual artifacts do not influence our results.

We compared in detail the morphology and fractional area covered by reversed-polarity Stokes~V profiles (RPPs) and three-lobed Stokes~V profiles (3LPs) in visible and IR spectral lines for sunspots close to disk center. The patches of RPPs have a roundish shape and are located in the outer penumbra. In the observation of visible spectral lines, they are about twice as frequent as in the IR observation. Regions with 3LPs appear roundish and are located in the middle and outer penumbra. In Fe~I~630.25~nm, however, their size and frequency is up to one order of magnitude larger than in Fe~I~1566.2~nm. In the data of the visible spectral lines, some patches appear elongated and are located in the center-side penumbra. This cannot be caused by projection effects and indicates true penumbral return flux. Such elongated features are absent from the IR data. From our observation, we found that magnetic fields of opposite polarity cover a significantly smaller fraction of the penumbra when observed in IR instead of visible lines.

We performed spectral inversion and forward synthesis on a typical 3LP in Fe~I~630.25~nm and Fe~I~1566.2~nm to verify that our algorithm is able to detect 3LPs in the Hinode-SP and GRIS data. We studied the center-to-limb variation of penumbral 3LPs and used a polynomial fit to extrapolate our results to $\Uptheta=0^{\circ}$. We found that at disk center, 13.5\% of the penumbra is covered with 3LPs for Fe~I~630.25~nm, but only 1.5\% for Fe~I~1566.2~nm. For a given heliocentric angle, the morphology of the patches of penumbral 3LPs is similar in all of the spectral lines. For increasing $\Uptheta$, their position shifts toward the limb-side penumbra, where they appear predominantly along the magnetic neutral line and follow the filamentary structure of dark penumbral filaments. These results demonstrate that the different findings in visible and IR data are not due to selection bias, but a systematic effect.

An explanation of the different numbers of RPP patches in the Hinode-SP and GRIS data seems possible within the framework of the following scenario: The downflows in the outer penumbra are typically associated with magnetic fields that are stronger than the penumbral background field
\citep{2013A&A...557A..25T, 
2013A&A...550A..97F}. 
Such a flux concentration or flux tube causes evacuation, and spectral lines are formed deeper within that tube than outside of it. As a result of pressure balance and the stratification with height of the photosphere, the flux tube narrows with depth. The area showing RPPs is therefor expected to be smaller for lines which form deeper, that is, for the IR lines.

Asymmetric Stokes V profiles are commonly explained using a multilayer atmosphere with a strong gradient or discontinuity of the atmospheric parameters, especially the velocity, along the LOS
\citep{1978A&A....64...67A, 
1983SoPh...87..221L}. 
Such a configuration cannot only reproduce the observed net circular polarization, but also causes 3LPs if unequal plasma flows and magnetic fields of opposite polarity are present within the atmosphere
\citep{2004A&A...422.1093B, 
2013A&A...550A..97F}. 
A scenario with strong plasma flows and magnetic fields of opposite polarity in the lower atmospheric layer would also explain the asymmetric Stokes I profiles with the strongly shifted line wing. 
If the lines observed by Hinode form in a similar atmospheric region as the IR lines of GRIS, then we should measure a similar amount of penumbral 3LPs. This is not the case. If the Hinode lines form in such a way that they sense the gradient of the atmospheric parameters, while the GRIS lines, which form in a deeper and narrower layer, are not sensitive to the gradient, then we should expect more RPPs in the near-infrared lines. We found that this is not the case. To reproduce our findings, an atmosphere with opposite-polarity fields and downflows only in the higher layer would be required, at least in the context of the simple model discussed above. Such an atmospheric configuration seems unphysical and cannot explain the strong line shift in the wing of Stokes I profiles.

It should be noted that the low formation height of IR lines does not necessarily mean that they are equally well-suited for studying the atmospheric conditions of all physical parameters in the deep photosphere. Using an analytical model for the shape of spectral lines, 
\citet{2005A&A...439..687C} 
showed that the response of Fe~I~1564.85~nm to velocity perturbation is weaker than of Fe~I~630.25~nm. The occurrence of a 3LP requires plasma velocities in the deep line-formation layers. If Fe~I~1564.85~nm is less sensitive to velocities, then the uncertainty in the determination of the third lobe in the Stokes V profile would increase and the number of detected 3LPs would decrease.

Unfortunately, it is not straightforward to apply these results to our investigation. The study of 
\citet{2005A&A...439..687C} 
investigated one atmospheric parameter at a time, while a 3LP requires combined plasma velocities and magnetic fields with various inclinations in different layers of the line-forming region of the atmosphere. Furthermore, the approximation of a spectral line with a Gaussian function encounters difficulties for asymmetric Stokes I profiles with strongly shifted line wings. Finally, the sensitivities of Fe~I~1565.29~nm and Fe~I~1566.20~nm to velocity perturbations were not studied by these authors. Especially the Fe~I~1566.20~nm line, which has a higher line-depth to line-width shape ratio, should be more sensitive to velocities.

To improve the observational data base, we plan to perform coordinated observations of sunspots with Hinode, the GREGOR infrared spectrograph, and the GREGOR Fabry Perot Interferometer
\citep[GFPI,][]{2012AN....333..880P, 
2013OptEn..52h1606P}. 
From such strictly simultaneous observations, information about the strength and inclination of the magnetic field at various atmospheric heights could be obtained. This would enable us to calculate numbers for the penumbral return flux throughout the atmosphere.

The amount of spatial stray light is another difference \mbox{between} the Hinode measurements and any ground-based observation. In the seeing-free environment of space, the spatial stray light is only influenced by the telescope and the instrument itself. Even then, a number of instruments display significant amounts of stray light going beyond the theoretical PSF, as deduced from a transit of Mercury for the Hinode Broadband Filter Imager
\citep{2009A&A...503..225W,
2009A&A...501L..19M}, 
from limb profiles for the Michelson Doppler Imager 
\citep{2007A&A...465..291M}, 
or from a Venus transit for the Helioseismic and Magnetic Imager
\citep{2014A&A...561A..22Y}. 
\citet{2008A&A...484L..17D} 
degraded magneto-hydrodynamic simulations with a theoretical PSF and found excellent agreement to observation of the spectropolarimeter onboard Hinode in terms of the root-mean-square contrast of the continuum intensity and its probability-density function 
\citep[see][for an application of this strategy to Sunrise data]{2014A&A...568A..13R}. 
\citet{2016A&A...Lagg} used this method to approximate the power outside the core of the PSF and to estimate the amount of spatial stray light in GRIS data. However, this technique does not allow determining the exact shape of the PSF, which is required for a deconvolution of the GRIS data prior to the analysis.

\citet{2016A&A...Borrero} 
approximated the PSF with two Gaussians. To obtain the required parameters, they used pinhole measurements and an average umbral Stokes I profile, where the magnetic field is oriented along the line of sight. An umbral Fe~I~1564.85~nm Stokes I profile should show only the $\upsigma$-components of the transition, while any indication of $\uppi$-component is ascribed to the influence of stray light. The drawback of this method is that the observations allow for a large number of parameters and do not yield a unique solution for the PSF.

In the future, we plan to synthesize the visible Hinode lines at 630.2~nm and the infrared lines of GREGOR around 1.55~$\upmu$m from a magnetohydrodynamic sunspot simulation 
\citep[see][]{2012ApJ...750...62R, 
2014A&A...572A..54B}. 
Degrading these synthetic data spatially and spectrally with a number of theoretical PSFs provides another possibility of studying the influence of stray light on the detectable amount of RPPs and 3LPs in the penumbra. The PSF that yields the best match between degraded synthetic data and GRIS observation could be used to compensate for the influence of stray light.


\begin{acknowledgements} 
We wish to thank J. C. del Toro Iniesta for valuable comments on the manuscript. The 1.5-meter GREGOR  solar telescope was built by a German consortium under the leadership of the Kiepenheuer Institut f\"ur Sonnenphysik in Freiburg with the Leibniz Institut f\"ur Astrophysik Potsdam, the Institut f\"ur Astrophysik G\"ottingen, and the Max-Planck Institut f\"ur Sonnensystemforschung in G\"ottingen as partners, and with contributions by the Instituto de Astrof\'sica de Canarias and the Astronomical Institute of the Academy of Sciences of the Czech Republic. The GRIS instrument was developed thanks to the support by the Spanish Ministry of Economy and Competitiveness through the project AYA2010-18029 (Solar Magnetism and Astrophysical Spectropolarimetry). Hinode is a Japanese mission developed and launched by ISAS/JAXA, with NAOJ as domestic partner and NASA and STFC (UK) as international partners. It is operated by these agencies in cooperation with ESA and NSC (Norway). NSO/Kitt Peak FTS data used here were produced by NSF/NOAO. This research has made use of NASA's Astrophysics Data System.
\end{acknowledgements}


\bibliographystyle{aa} 
\bibliography{references,references2}

\begin{thebibliography}{80}
\expandafter\ifx\csname natexlab\endcsname\relax\def\natexlab#1{#1}\fi

\bibitem[{{Allende Prieto} {et~al.}(2004){Allende Prieto}, {Asplund}, \&
  {Fabiani Bendicho}}]{2004A&A...423.1109A}
{Allende Prieto}, C., {Asplund}, M., \& {Fabiani Bendicho}, P. 2004, \aap, 423,
  1109

\bibitem[{{Auer} \& {Heasley}(1978)}]{1978A&A....64...67A}
{Auer}, L.~H. \& {Heasley}, J.~N. 1978, \aap, 64, 67

\bibitem[{{Bellot Rubio} {et~al.}(2004){Bellot Rubio}, {Balthasar}, \&
  {Collados}}]{2004A&A...427..319B}
{Bellot Rubio}, L.~R., {Balthasar}, H., \& {Collados}, M. 2004, \aap, 427, 319

\bibitem[{{Bellot Rubio} {et~al.}(2003){Bellot Rubio}, {Balthasar}, {Collados},
  \& {Schlichenmaier}}]{2003A&A...403L..47B}
{Bellot Rubio}, L.~R., {Balthasar}, H., {Collados}, M., \& {Schlichenmaier}, R.
  2003, \aap, 403, L47

\bibitem[{{Berkefeld } {et~al.}(2012){Berkefeld }, {Schmidt}, {Soltau}, {von
  der L{\"u}he}, \& {Heidecke}}]{2012AN....333..863B}
{Berkefeld }, T., {Schmidt}, D., {Soltau}, D., {von der L{\"u}he}, O., \&
  {Heidecke}, F. 2012, Astronomische Nachrichten, 333, 863

\bibitem[{Borrero {et~al.}(2016)Borrero, Asensio~Ramos, Collados,
  Schlichenmaier, Balthasar, Franz, Rezaei, Orozco~Su\'arez, \&
  et~al.}]{2016A&A...Borrero}
Borrero, J., Asensio~Ramos, A., Collados, M., {et~al.} 2016, \aap, \it{this
  issue}\rm

\bibitem[{{Borrero} \& {Ichimoto}(2011)}]{2011LRSP....8....4B}
{Borrero}, J.~M. \& {Ichimoto}, K. 2011, Living Reviews in Solar Physics, 8
  [\eprint[arXiv]{1109.4412}]

\bibitem[{{Borrero} {et~al.}(2005){Borrero}, {Lagg}, {Solanki}, \&
  {Collados}}]{2005A&A...436..333B}
{Borrero}, J.~M., {Lagg}, A., {Solanki}, S.~K., \& {Collados}, M. 2005, \aap,
  436, 333

\bibitem[{{Borrero} {et~al.}(2014){Borrero}, {Lites}, {Lagg}, {Rezaei}, \&
  {Rempel}}]{2014A&A...572A..54B}
{Borrero}, J.~M., {Lites}, B.~W., {Lagg}, A., {Rezaei}, R., \& {Rempel}, M.
  2014, \aap, 572, A54

\bibitem[{{Borrero} {et~al.}(2004){Borrero}, {Solanki}, {Bellot Rubio}, {Lagg},
  \& {Mathew}}]{2004A&A...422.1093B}
{Borrero}, J.~M., {Solanki}, S.~K., {Bellot Rubio}, L.~R., {Lagg}, A., \&
  {Mathew}, S.~K. 2004, \aap, 422, 1093

\bibitem[{{Cabrera Solana} {et~al.}(2007){Cabrera Solana}, {Bellot Rubio},
  {Beck}, \& {Del Toro Iniesta}}]{2007A&A...475.1067C}
{Cabrera Solana}, D., {Bellot Rubio}, L.~R., {Beck}, C., \& {Del Toro Iniesta},
  J.~C. 2007, \aap, 475, 1067

\bibitem[{{Cabrera Solana} {et~al.}(2005){Cabrera Solana}, {Bellot Rubio}, \&
  {del Toro Iniesta}}]{2005A&A...439..687C}
{Cabrera Solana}, D., {Bellot Rubio}, L.~R., \& {del Toro Iniesta}, J.~C. 2005,
  \aap, 439, 687

\bibitem[{{Casini} {et~al.}(2012){Casini}, {de Wijn}, \&
  {Judge}}]{2012ApJ...757...45C}
{Casini}, R., {de Wijn}, A.~G., \& {Judge}, P.~G. 2012, \apj, 757, 45

\bibitem[{{Collados}(1999)}]{1999ASPC..184....3C}
{Collados}, M. 1999, in Astronomical Society of the Pacific Conference Series,
  Vol. 184, Third Advances in Solar Physics Euroconference: Magnetic Fields and
  Oscillations, ed. B.~{Schmieder}, A.~{Hofmann}, \& J.~{Staude}, 3--22

\bibitem[{{Collados} {et~al.}(2012){Collados}, {L{\'o}pez}, {P{\'a}ez},
  {Hern{\'a}ndez}, {Reyes}, {Calcines}, {Ballesteros}, {D{\'{\i}}az}, {Denker},
  {Lagg}, {Schlichenmaier}, {Schmidt}, {Solanki}, {Strassmeier}, {von der
  L{\"u}he}, \& {Volkmer}}]{2012AN....333..872C}
{Collados}, M., {L{\'o}pez}, R., {P{\'a}ez}, E., {et~al.} 2012, Astronomische
  Nachrichten, 333, 872

\bibitem[{{Danilovic} {et~al.}(2008){Danilovic}, {Gandorfer}, {Lagg},
  {Sch{\"u}ssler}, {Solanki}, {V{\"o}gler}, {Katsukawa}, \&
  {Tsuneta}}]{2008A&A...484L..17D}
{Danilovic}, S., {Gandorfer}, A., {Lagg}, A., {et~al.} 2008, \aap, 484, L17

\bibitem[{{del Toro Iniesta}(2003)}]{2003isp..book.....D}
{del Toro Iniesta}, J.~C. 2003, {Introduction to Spectropolarimetry}
  ({Cambridge University Press, Cambridge, UK}), 244

\bibitem[{{del Toro Iniesta} {et~al.}(2001){del Toro Iniesta}, {Bellot Rubio},
  \& {Collados}}]{2001ApJ...549L.139D}
{del Toro Iniesta}, J.~C., {Bellot Rubio}, L.~R., \& {Collados}, M. 2001,
  \apjl, 549, L139

\bibitem[{{Denker} {et~al.}(2012){Denker}, {von der L{\"u}he}, {Feller},
  {Arlt}, {Balthasar}, {Bauer}, {Bello Gonz{\'a}lez}, {Berkefeld}, {Caligari},
  {Collados}, {Fischer}, {Granzer}, {Hahn}, {Halbgewachs}, {Heidecke},
  {Hofmann}, {Kentischer}, Klvana, {Kneer}, {Lagg}, {Nicklas}, {Popow},
  {Puschmann}, {Rendtel}, {Schmidt}, {Schmidt}, {Sobotka}, {Solanki}, {Soltau},
  {Staude}, {Strassmeier}, {Volkmer}, {Waldmann}, {Wiehr}, {Wittmann}, \&
  {Woche}}]{2012AN....333..810D}
{Denker}, C., {von der L{\"u}he}, O., {Feller}, A., {et~al.} 2012,
  Astronomische Nachrichten, 333, 810

\bibitem[{{Franz}(2011)}]{2011PhDT.......137F}
{Franz}, M. 2011, PhD thesis, University of Freiburg, 2011, 176 pp.

\bibitem[{{Franz}(2012)}]{2012AN....333.1009F}
{Franz}, M. 2012, Astronomische Nachrichten, 333, 1009

\bibitem[{{Franz} \& {Schlichenmaier}(2009)}]{2009A&A...508.1453F}
{Franz}, M. \& {Schlichenmaier}, R. 2009, \aap, 508, 1453

\bibitem[{{Franz} \& {Schlichenmaier}(2013)}]{2013A&A...550A..97F}
{Franz}, M. \& {Schlichenmaier}, R. 2013, \aap, 550, A97

\bibitem[{{Giampapa}(1982)}]{1982Giampapa}
{Giampapa}, M. 1982, {Spectrum in the NSO-FTS archive.} ({N.S.O. Tech. Report
  1982/05/16\#05})

\bibitem[{{Giampapa}(1983)}]{1983Giampapa}
{Giampapa}, M. 1983, {Spectrum in the NSO-FTS archive.} ({N.S.O. Tech. Report
  1983/06/16\#01})

\bibitem[{{Grossmann-Doerth} {et~al.}(1989){Grossmann-Doerth}, {Schuessler}, \&
  {Solanki}}]{1989A&A...221..338G}
{Grossmann-Doerth}, U., {Schuessler}, M., \& {Solanki}, S.~K. 1989, \aap, 221,
  338

\bibitem[{{Hofmann} {et~al.}(2012){Hofmann}, {Arlt}, {Balthasar}, {Bauer},
  {Bittner}, {Paschke}, {Popow}, {Rendtel}, {Soltau}, \&
  {Waldmann}}]{2012AN....333..854H}
{Hofmann}, A., {Arlt}, K., {Balthasar}, H., {et~al.} 2012, Astronomische
  Nachrichten, 333, 854

\bibitem[{{Ichimoto} {et~al.}(2007){Ichimoto}, {Shine}, {Lites}, {Kubo},
  {Shimizu}, {Suematsu}, {Tsuneta}, {Katsukawa}, {Tarbell}, {Title}, {Nagata},
  {Yokoyama}, \& {Shimojo}}]{2007PASJ...59S.593I}
{Ichimoto}, K., {Shine}, R.~A., {Lites}, B., {et~al.} 2007, \pasj, 59, S593

\bibitem[{{Illing} {et~al.}(1975){Illing}, {Landman}, \&
  {Mickey}}]{1975A&A....41..183I}
{Illing}, R.~M.~E., {Landman}, D.~A., \& {Mickey}, D.~L. 1975, \aap, 41, 183

\bibitem[{{Katsukawa} \& {Jur{\v c}{\'a}k}(2010)}]{2010A&A...524A..20K}
{Katsukawa}, Y. \& {Jur{\v c}{\'a}k}, J. 2010, \aap, 524, A20

\bibitem[{Lagg {et~al.}(2016)Lagg, Sami K.~Solanki, Doerr, Riethm{\"}uller,
  Schlichenmaier, Orozco~Su{\'}árez, Mart{\'}inez~Gonz{\'}alez, \&
  et~al.}]{2016A&A...Lagg}
Lagg, A., Sami K.~Solanki, S.~K., Doerr, H.~P., {et~al.} 2016, \aap, \it{this
  issue}\rm

\bibitem[{{Landi Degl'Innocenti} \& {Landolfi}(1983)}]{1983SoPh...87..221L}
{Landi Degl'Innocenti}, E. \& {Landolfi}, M. 1983, \solphys, 87, 221

\bibitem[{{Langhans} {et~al.}(2005){Langhans}, {Scharmer}, {Kiselman},
  {L{\"o}fdahl}, \& {Berger}}]{2005A&A...436.1087L}
{Langhans}, K., {Scharmer}, G.~B., {Kiselman}, D., {L{\"o}fdahl}, M.~G., \&
  {Berger}, T.~E. 2005, \aap, 436, 1087

\bibitem[{{Lites} {et~al.}(2013){Lites}, {Akin}, {Card}, {Cruz}, {Duncan},
  {Edwards}, {Elmore}, {Hoffmann}, {Katsukawa}, {Katz}, {Kubo}, {Ichimoto},
  {Shimizu}, {Shine}, {Streander}, {Suematsu}, {Tarbell}, {Title}, \&
  {Tsuneta}}]{2013SoPh..283..579L}
{Lites}, B.~W., {Akin}, D.~L., {Card}, G., {et~al.} 2013, \solphys, 283, 579

\bibitem[{{Lites} {et~al.}(1993){Lites}, {Elmore}, {Seagraves}, \&
  {Skumanich}}]{1993ApJ...418..928L}
{Lites}, B.~W., {Elmore}, D.~F., {Seagraves}, P., \& {Skumanich}, A.~P. 1993,
  \apj, 418, 928

\bibitem[{{Lites} {et~al.}(2001){Lites}, {Elmore}, \&
  {Streander}}]{2001ASPC..236...33L}
{Lites}, B.~W., {Elmore}, D.~F., \& {Streander}, K.~V. 2001, in Astronomical
  Society of the Pacific Conference Series, Vol. 236, Advanced Solar
  Polarimetry -- Theory, Observation, and Instrumentation, ed. M.~{Sigwarth},
  33

\bibitem[{{Lites} \& {Ichimoto}(2013)}]{2013SoPh..283..601L}
{Lites}, B.~W. \& {Ichimoto}, K. 2013, \solphys, 283, 601

\bibitem[{{Livingston}(1990)}]{1990Livingston}
{Livingston}, W. 1990, {Spectrum in the NSO-FTS archive.} ({N.S.O. Tech. Report
  1990/12/18\#06})

\bibitem[{{Livingston} \& {Wallace}(1991)}]{1991aass.book.....L}
{Livingston}, W. \& {Wallace}, L. 1991, {An atlas of the solar spectrum in the
  infrared from 1850 to 9000 cm-1 (1.1 to 5.4 micrometer)} ({N.S.O. Tech.
  Report \#91-001})

\bibitem[{{Maltby}(1964)}]{1964ApNr....8..205M}
{Maltby}, P. 1964, Astrophysica Norvegica, 8, 205

\bibitem[{{M{\'a}rtinez Pillet} {et~al.}(1999){M{\'a}rtinez Pillet},
  {Collados}, {S{\'a}nchez Almeida}, {Gonz{\'a}lez}, {Cruz-Lopez}, {Manescau},
  {Joven}, {Paez}, {Diaz}, {Feeney}, {S{\'a}nchez}, {Scharmer}, \&
  {Soltau}}]{1999ASPC..183..264M}
{M{\'a}rtinez Pillet}, V., {Collados}, M., {S{\'a}nchez Almeida}, J., {et~al.}
  1999, in Astronomical Society of the Pacific Conference Series, Vol. 183,
  High Resolution Solar Physics: Theory, Observations, and Techniques, ed.
  T.~R. {Rimmele}, K.~S. {Balasubramaniam}, \& R.~R. {Radick}, 264

\bibitem[{{Mathew} {et~al.}(2007){Mathew}, {Mart{\'{\i}}nez Pillet}, {Solanki},
  \& {Krivova}}]{2007A&A...465..291M}
{Mathew}, S.~K., {Mart{\'{\i}}nez Pillet}, V., {Solanki}, S.~K., \& {Krivova},
  N.~A. 2007, \aap, 465, 291

\bibitem[{{Mathew} {et~al.}(2009){Mathew}, {Zakharov}, \&
  {Solanki}}]{2009A&A...501L..19M}
{Mathew}, S.~K., {Zakharov}, V., \& {Solanki}, S.~K. 2009, \aap, 501, L19

\bibitem[{{Mihalas}(1978)}]{1978stat.book.....M}
{Mihalas}, D. 1978, {Stellar atmospheres /2nd edition/} (San Francisco,
  W.~H.~Freeman and Co., 1978.~650 p.)

\bibitem[{{M{\"u}ller} {et~al.}(2002){M{\"u}ller}, {Schlichenmaier}, {Steiner},
  \& {Stix}}]{2002A&A...393..305M}
{M{\"u}ller}, D.~A.~N., {Schlichenmaier}, R., {Steiner}, O., \& {Stix}, M.
  2002, \aap, 393, 305

\bibitem[{{Nave} {et~al.}(1994){Nave}, {Johansson}, {Learner}, {Thorne}, \&
  {Brault}}]{1994ApJS...94..221N}
{Nave}, G., {Johansson}, S., {Learner}, R.~C.~M., {Thorne}, A.~P., \& {Brault},
  J.~W. 1994, \apjs, 94, 221

\bibitem[{{Penn}(2014)}]{2014LRSP...11....2P}
{Penn}, M.~J. 2014, Living Reviews in Solar Physics, 11

\bibitem[{{Pesnell} {et~al.}(2012){Pesnell}, {Thompson}, \&
  {Chamberlin}}]{2012SoPh..275....3P}
{Pesnell}, W.~D., {Thompson}, B.~J., \& {Chamberlin}, P.~C. 2012, \solphys,
  275, 3

\bibitem[{{Puschmann} {et~al.}(2013){Puschmann}, {Denker}, {Balthasar},
  {Louis}, {Popow}, {Woche}, {Beck}, {Seelemann}, \&
  {Volkmer}}]{2013OptEn..52h1606P}
{Puschmann}, K.~G., {Denker}, C., {Balthasar}, H., {et~al.} 2013, Optical
  Engineering, 52, 081606

\bibitem[{{Puschmann} {et~al.}(2012){Puschmann}, {Denker}, {Kneer}, {Al
  Erdogan}, {Balthasar}, {Bauer}, {Beck}, {Bello Gonz{\'a}lez}, {Collados},
  {Hahn}, {Hirzberger}, {Hofmann}, {Louis}, {Nicklas}, {Okunev},
  {Mart{\'{\i}}nez Pillet}, {Popow}, {Seelemann}, {Volkmer}, {Wittmann}, \&
  {Woche}}]{2012AN....333..880P}
{Puschmann}, K.~G., {Denker}, C., {Kneer}, F., {et~al.} 2012, Astronomische
  Nachrichten, 333, 880

\bibitem[{{Ramsauer} {et~al.}(1995){Ramsauer}, {Solanki}, \&
  {Biemont}}]{1995A&AS..113...71R}
{Ramsauer}, J., {Solanki}, S.~K., \& {Biemont}, E. 1995, \aaps, 113, 71

\bibitem[{{Rempel}(2012)}]{2012ApJ...750...62R}
{Rempel}, M. 2012, \apj, 750, 62

\bibitem[{{Rempel} \& {Schlichenmaier}(2011)}]{2011LRSP....8....3R}
{Rempel}, M. \& {Schlichenmaier}, R. 2011, Living Reviews in Solar Physics, 8

\bibitem[{{Riethm{\"u}ller} {et~al.}(2014){Riethm{\"u}ller}, {Solanki},
  {Berdyugina}, {Sch{\"u}ssler}, {Mart{\'{\i}}nez Pillet}, {Feller},
  {Gandorfer}, \& {Hirzberger}}]{2014A&A...568A..13R}
{Riethm{\"u}ller}, T.~L., {Solanki}, S.~K., {Berdyugina}, S.~V., {et~al.} 2014,
  \aap, 568, A13

\bibitem[{{Rojo} \& {Harrington}(2006)}]{2006ApJ...649..553R}
{Rojo}, P.~M. \& {Harrington}, J. 2006, \apj, 649, 553

\bibitem[{{Ruiz Cobo} \& {Asensio Ramos}(2013)}]{2013A&A...549L...4R}
{Ruiz Cobo}, B. \& {Asensio Ramos}, A. 2013, \aap, 549, L4

\bibitem[{{Ruiz Cobo} \& {del Toro Iniesta}(1992)}]{1992ApJ...398..375R}
{Ruiz Cobo}, B. \& {del Toro Iniesta}, J.~C. 1992, \apj, 398, 375

\bibitem[{{S{\'a}nchez Almeida} \& {Ichimoto}(2009)}]{2009A&A...508..963S}
{S{\'a}nchez Almeida}, J. \& {Ichimoto}, K. 2009, \aap, 508, 963

\bibitem[{{Sanchez Almeida} \& {Lites}(1992)}]{1992ApJ...398..359S}
{Sanchez Almeida}, J. \& {Lites}, B.~W. 1992, \apj, 398, 359

\bibitem[{{Scharmer} {et~al.}(2013){Scharmer}, {de la Cruz Rodriguez},
  {S{\"u}tterlin}, \& {Henriques}}]{2013A&A...553A..63S}
{Scharmer}, G.~B., {de la Cruz Rodriguez}, J., {S{\"u}tterlin}, P., \&
  {Henriques}, V.~M.~J. 2013, \aap, 553, A63

\bibitem[{{Schmidt} {et~al.}(2012){Schmidt}, {von der L{\"u}he}, {Volkmer},
  {Denker}, {Solanki}, {Balthasar}, {Bello Gonzalez}, {Berkefeld}, {Collados},
  {Fischer}, {Halbgewachs}, {Heidecke}, {Hofmann}, {Kneer}, {Lagg}, {Nicklas},
  {Popow}, {Puschmann}, {Schmidt}, {Sigwarth}, {Sobotka}, {Soltau}, {Staude},
  {Strassmeier}, \& {Waldmann }}]{2012AN....333..796S}
{Schmidt}, W., {von der L{\"u}he}, O., {Volkmer}, R., {et~al.} 2012,
  Astronomische Nachrichten, 333, 796

\bibitem[{{Schou} {et~al.}(2012){Schou}, {Scherrer}, {Bush}, {Wachter},
  {Couvidat}, {Rabello-Soares}, {Bogart}, {Hoeksema}, {Liu}, {Duvall}, {Akin},
  {Allard}, {Miles}, {Rairden}, {Shine}, {Tarbell}, {Title}, {Wolfson},
  {Elmore}, {Norton}, \& {Tomczyk}}]{2012SoPh..275..229S}
{Schou}, J., {Scherrer}, P.~H., {Bush}, R.~I., {et~al.} 2012, \solphys, 275,
  229

\bibitem[{{Semel}(2003)}]{2003A&A...401....1S}
{Semel}, M. 2003, \aap, 401, 1

\bibitem[{{Solanki}(2003)}]{2003A&ARv..11..153S}
{Solanki}, S.~K. 2003, \aapr, 11, 153

\bibitem[{{Solanki} \& {Montavon}(1993)}]{1993A&A...275..283S}
{Solanki}, S.~K. \& {Montavon}, C.~A.~P. 1993, \aap, 275, 283

\bibitem[{{Solanki} {et~al.}(1994){Solanki}, {Montavon}, \&
  {Livingston}}]{1994A&A...283..221S}
{Solanki}, S.~K., {Montavon}, C.~A.~P., \& {Livingston}, W. 1994, \aap, 283,
  221

\bibitem[{{Solanki} {et~al.}(1987){Solanki}, {Pantellini}, \&
  {Stenflo}}]{1986SoPh..107...57S}
{Solanki}, S.~K., {Pantellini}, F.~G.~E., \& {Stenflo}, J.~O. 1987, \solphys,
  107, 57

\bibitem[{{Solanki} {et~al.}(1992){Solanki}, {Rueedi}, \&
  {Livingston}}]{1992A&A...263..312S}
{Solanki}, S.~K., {Rueedi}, I.~K., \& {Livingston}, W. 1992, \aap, 263, 312

\bibitem[{{Spruit} \& {Scharmer}(2006)}]{2006A&A...447..343S}
{Spruit}, H.~C. \& {Scharmer}, G.~B. 2006, \aap, 447, 343

\bibitem[{{Tiwari} {et~al.}(2013){Tiwari}, {van Noort}, {Lagg}, \&
  {Solanki}}]{2013A&A...557A..25T}
{Tiwari}, S.~K., {van Noort}, M., {Lagg}, A., \& {Solanki}, S.~K. 2013, \aap,
  557, A25

\bibitem[{{Tiwari} {et~al.}(2015){Tiwari}, {van Noort}, {Solanki}, \&
  {Lagg}}]{2015A&A...583A.119T}
{Tiwari}, S.~K., {van Noort}, M., {Solanki}, S.~K., \& {Lagg}, A. 2015, \aap,
  583, A119

\bibitem[{{Tsuneta} {et~al.}(2008){Tsuneta}, {Ichimoto}, {Katsukawa}, {Nagata},
  {Otsubo}, {Shimizu}, {Suematsu}, {Nakagiri}, {Noguchi}, {Tarbell}, {Title},
  {Shine}, {Rosenberg}, {Hoffmann}, {Jurcevich}, {Kushner}, {Levay}, {Lites},
  {Elmore}, {Matsushita}, {Kawaguchi}, {Saito}, {Mikami}, {Hill}, \&
  {Owens}}]{2008SoPh..249..167T}
{Tsuneta}, S., {Ichimoto}, K., {Katsukawa}, Y., {et~al.} 2008, \solphys, 249,
  167

\bibitem[{{Uns{\"o}ld} \& {Baschek}(2002)}]{2002neko.book.....U}
{Uns{\"o}ld}, A. \& {Baschek}, B. 2002, {Der neue Kosmos. Einf{\"u}hrung in die
  Astronomie und Astrophysik} ({Springer, Berlin, Germany}), 575

\bibitem[{{Volkmer} {et~al.}(2012){Volkmer}, {Eisentr{\"a}ger}, {Emde},
  {Fischer}, {von der L{\"u}he}, {Nicklas}, {Soltau}, {Schmidt}, \&
  {Weis}}]{2012AN....333..816V}
{Volkmer}, R., {Eisentr{\"a}ger}, P., {Emde}, P., {et~al.} 2012, Astronomische
  Nachrichten, 333, 816

\bibitem[{{von der L{\"u}he} {et~al.}(2001){von der L{\"u}he}, {Schmidt},
  {Soltau}, {Berkefeld}, {Kneer}, \& {Staude}}]{2001AN....322..353V}
{von der L{\"u}he}, O., {Schmidt}, W., {Soltau}, D., {et~al.} 2001,
  Astronomische Nachrichten, 322, 353

\bibitem[{{Wallace} {et~al.}(2001){Wallace}, {Hinkle}, \&
  {Livingston}}]{2001sus..book.....W}
{Wallace}, L., {Hinkle}, K., \& {Livingston}, W. 2001, {Sunspot umbral spectra
  in the region 4000 to 8640 cm(-1) (1.16 to 2.50 [microns])} ({N.S.O. Tech.
  Report \#01-001})

\bibitem[{{Wedemeyer-B{\"o}hm} \& {Rouppe van der
  Voort}(2009)}]{2009A&A...503..225W}
{Wedemeyer-B{\"o}hm}, S. \& {Rouppe van der Voort}, L. 2009, \aap, 503, 225

\bibitem[{{Westendorp Plaza} {et~al.}(1997){Westendorp Plaza}, {del Toro
  Iniesta}, {Ruiz Cobo}, {Martinez Pillet}, {Lites}, \&
  {Skumanich}}]{1997Natur.389...47W}
{Westendorp Plaza}, C., {del Toro Iniesta}, J.~C., {Ruiz Cobo}, B., {et~al.}
  1997, \nat, 389, 47

\bibitem[{{Westendorp Plaza} {et~al.}(2001){Westendorp Plaza}, {del Toro
  Iniesta}, {Ruiz Cobo}, {Mart{\'{\i}}nez Pillet}, {Lites}, \&
  {Skumanich}}]{2001ApJ...547.1130W}
{Westendorp Plaza}, C., {del Toro Iniesta}, J.~C., {Ruiz Cobo}, B., {et~al.}
  2001, \apj, 547, 1130

\bibitem[{{Yeo} {et~al.}(2014){Yeo}, {Feller}, {Solanki}, {Couvidat},
  {Danilovic}, \& {Krivova}}]{2014A&A...561A..22Y}
{Yeo}, K.~L., {Feller}, A., {Solanki}, S.~K., {et~al.} 2014, \aap, 561, A22

\end{thebibliography}


\clearpage
\begin{appendix}
\section{List of sunspots}
\label{sec:app01}

\begin{table}[h!]
\begin{center}
	\caption{Details of Hinode data.}
	\begin{tabular}{ccccc}
		\hline
		\hline
		\\[-2ex]
		{NOAA} & {Date of} & {heliocentric} & {1$\sigmaup$ noise} \\
		{active region} & {observation} & {angle [cos$(\Theta$)]} & {[$10^{-4}\cdot\rm{I}_{\rm{c}}$]} \\
		\\[-2ex]
		\hline		
		\\[-2ex]
		{\#01 10923} & {Nov 10$^{\rm{th}}$ 2006} & {0.585 - 0.680} & {11.7}\\	
		{\#02 10923} & {Nov 11$^{\rm{th}}$ 2006} & {0.741 - 0.818} & {11.3}\\
		{\#03 10923} & {Nov 12$^{\rm{th}}$ 2006} & {0.834 - 0.887} & {10.8}\\
		{\#04 10923} & {Nov 13$^{\rm{th}}$ 2006} & {0.957 - 0.983} & {10.6}\\
		{\#05 10923} & {Nov 14$^{\rm{th}}$ 2006} & {0.982 - 0.995} & {10.5}\\
	 	{\#06 10923} & {Nov 14$^{\rm{th}}$ 2006} & {0.979 - 0.994} & {10.6}\\
		{\#07 10923} & {Nov 15$^{\rm{th}}$ 2006} & {0.944 - 0.970} & {10.7}\\
		{\#08 10923} & {Nov 15$^{\rm{th}}$ 2006} & {0.934 - 0.961} & {10.8}\\
		{\#09 10923} & {Nov 16$^{\rm{th}}$ 2006} & {0.838 - 0.883} & {10.6}\\
		{\#10 10923} & {Nov 16$^{\rm{th}}$ 2006} & {0.798 - 0.847} & {10.7}\\
		{\#11 10923} & {Nov 18$^{\rm{th}}$ 2006} & {0.594 - 0.665} & {12.1}\\
		{\#12 10923} & {Nov 18$^{\rm{th}}$ 2006} & {0.455 - 0.536} & {12.9}\\
		{\#13 10923} & {Nov 19$^{\rm{th}}$ 2006} & {0.360 - 0.443} & {13.8}\\
		{\#14 10923} & {Nov 20$^{\rm{th}}$ 2006} & {0.208 - 0.301} & {15.3}\\
	
		{\#15 10930} & {Dec 6$^{\rm{th}}$ 2006} & {0.368 - 0.441} & {12.5}\\
		{\#16 10930} & {Dec 8$^{\rm{th}}$ 2006} & {0.698 - 0.750} & {10.8}\\
		{\#17 10930} & {Dec 10$^{\rm{th}}$ 2006} & {0.961 - 0.979} & {10.4}\\
		{\#18 10930} & {Dec 11$^{\rm{th}}$ 2006} & {0.994 - 0.999} & {10.3}\\
		{\#19 10930} & {Dec 12$^{\rm{th}}$ 2006} & {0.960 - 0.979} & {10.6}\\
		{\#20 10930} & {Dec 13$^{\rm{th}}$ 2006} & {0.866 - 0.903} & {11.0}\\
		{\#21 10930} & {Dec 14$^{\rm{th}}$ 2006} & {0.736 - 0.791} & {11.6}\\
		{\#22 10930} & {Dec 15$^{\rm{th}}$ 2006} & {0.564 - 0.625} & {12.3}\\
		{\#23 10930} & {Dec 16$^{\rm{th}}$ 2006} & {0.363 - 0.434} & {13.6}\\
		
		{\#24 10933} & {Jan 04$^{\rm{th}}$ 2007} & {0.952 - 0.968} & {10.2}\\
		{\#25 10933} & {Jan 05$^{\rm{th}}$ 2007} & {0.996 - 1.000} & {9.8}\\
		{\#26 10933} & {Jan 06$^{\rm{th}}$ 2007} & {0.974 - 0.987} & {10.3}\\
		{\#27 10933} & {Jan 07$^{\rm{th}}$ 2007} & {0.903 - 0.925} & {11.0}\\
		{\#28 10933} & {Jan 08$^{\rm{th}}$ 2007} & {0.809 - 0.838} & {11.0}\\
		{\#29 10933} & {Jan 09$^{\rm{th}}$ 2007} & {0.652 - 0.695} & {11.3}\\
		
		{\#30 10944} & {Feb 28$^{\rm{th}}$ 2007} & {0.999 - 1.000} & {10.0}\\
	    	{\#31 10944} & {Mar 01$^{\rm{st}}$ 2007} & {0.988 - 0.993} & {10.1}\\
	     	{\#32 10944} & {Mar 01$^{\rm{st}}$ 2007} & {0.976 - 0.984} & {10.3}\\
	     	{\#33 10944} & {Mar 01$^{\rm{st}}$ 2007} & {0.952 - 0.963} & {10.1}\\
	     	{\#34 10944} & {Mar 03$^{\rm{rd}}$ 2007} & {0.849 - 0.870} & {10.6}\\
	     	{\#35 10944} & {Mar 04$^{\rm{th}}$ 2007} & {0.708 - 0.735} & {10.6}\\
		
		{\#36 10953} & {Apr 30$^{\rm{th}}$ 2007} & {0.966 - 0.981} & {10.3}\\
		{\#37 10953} & {May 01$^{\rm{st}}$ 2007} & {0.986 - 0.997} & {8.3}\\
		{\#38 10953} & {May 02$^{\rm{nd}}$ 2007} & {0.967 - 0.981} & {10.6}\\
		{\#39 10953} & {May 03$^{\rm{rd}}$ 2007} & {0.895 - 0.923} & {10.8}\\
		{\#40 10953} & {May 04$^{\rm{th}}$ 2007} & {0.769 - 0.813} & {11.2}\\
		
		{\#41 10960} & {Jun 10$^{\rm{th}}$ 2007} & {0.786 - 0.808} & {8.8}\\
		{\#42 10960} & {Jun 11$^{\rm{th}}$ 2007} & {0.606 - 0.632} & {10.0}\\
		{\#43 10960} & {Jun 12$^{\rm{th}}$ 2007} & {0.467 - 0.496} & {9.8}\\

		{\#44 10969} & {Aug 27$^{\rm{th}}$ 2007} & {0.973 - 0.980} & {9.5}\\
		{\#45 10969} & {Aug 28$^{\rm{th}}$ 2007} & {0.948 - 0.960} & {10.4}\\	
		{\#46 10969} & {Aug 31$^{\rm{st}}$ 2007} & {0.590 - 0.612} & {9.9}\\
		\hline
		\end{tabular}	
	\label{Tab_A1}
\end{center}
\end{table}

\begin{table}[h!]
\begin{center}
	\caption{Details of GRIS data.}
	\begin{tabular}{cccc}
		\hline
		\hline
		\\[-2ex]
		{NOAA} & {Date of} & {heliocentric} & {1$\sigmaup$ noise} \\
		{active region} & {observation} & {angle [cos$(\Theta$)]} & {[$10^{-4}\cdot\rm{I}_{\rm{c}}$]} \\
		\\[-2ex]
		\hline		
		\\[-2ex]	
		
		{\#47 12045} & {Apr 26$^{\rm{th}}$ 2014} & {0.934 - 0.938} & {7.8}\\
		{\#48 12045} & {Apr 26$^{\rm{th}}$ 2014} & {0.934 - 0.938} & {11.2}\\
		{\#49 12045} & {Apr 26$^{\rm{th}}$ 2014} & {0.934 - 0.938} & {7.2}\\
		{\#50 12045} & {Apr 27$^{\rm{th}}$ 2014} & {0.889 - 0.894} & {8.9}\\
		{\#51 12045} & {Apr 27$^{\rm{th}}$ 2014} & {0.886 - 0.892} & {8.8}\\
		{\#52 12045} & {Apr 27$^{\rm{th}}$ 2014} & {0.886 - 0.891} & {6.7}\\
		
		{\#53 12049\footnotemark[1]} & {Apr 27$^{\rm{th}}$ 2014} & {0.250 - 0.266} & {7.2}\\
		{\#54 12049\footnotemark[1]} & {Apr 28$^{\rm{th}}$ 2014} & {0.403 - 0.421} & {7.4}\\
		{\#55 12049\footnotemark[1]} & {Apr 29$^{\rm{th}}$ 2014} & {0.600 - 0.615} & {6.9}\\
		{\#56 12049\footnotemark[1]} & {Apr 30$^{\rm{th}}$ 2014} & {0.767 - 0.781} & {7.0}\\
		{\#57 12049\footnotemark[1]} & {May 03$^{\rm{rd}}$ 2014} & {0.994 - 0.997} & {10.0}\\
		{\#58 12049\footnotemark[1]} & {May 03$^{\rm{rd}}$ 2014} & {0.995 - 0.997} & {10.3}\\
		{\#59 12049\footnotemark[1]} & {May 03$^{\rm{rd}}$ 2014} & {0.993 - 0.996} & {9.8}\\
		{\#60 12049\footnotemark[1]} & {May 03$^{\rm{rd}}$ 2014} & {0.992 - 0.995} & {7.8}\\
		{\#61 12049\footnotemark[1]} & {May 03$^{\rm{rd}}$ 2014} & {0.988 - 0.992} & {10.5}\\
		{\#62 12049\footnotemark[1]} & {May 05$^{\rm{th}}$ 2014} & {0.885 - 0.872} & {7.5}\\

		{\#63 12049\footnotemark[2]} & {May 03$^{\rm{rd}}$ 2014} & {0.997 - 0.998} & {10.1}\\
		{\#64 12049\footnotemark[2]} & {May 03$^{\rm{rd}}$ 2014} & {0.997 - 0.998} & {10.2}\\	
		{\#65 12049\footnotemark[2]} & {May 03$^{\rm{rd}}$ 2014} & {0.999 - 0.999} & {10.1}\\
		{\#66 12049\footnotemark[2]} & {May 03$^{\rm{rd}}$ 2014} & {0.999 - 0.999} & {9.9}\\
		{\#67 12049\footnotemark[2]} & {May 03$^{\rm{rd}}$ 2014} & {0.999 - 0.999} & {10.2}\\

		{\#68 12055} & {May 05$^{\rm{th}}$ 2014} & {0.289 - 0.338} & {10.4}\\

		{\#69 12104} & {Jul 03$^{\rm{rd}}$ 2014} & {0.889 - 0.915} & {8.3}\\	
		
		{\#70 12107} & {Jul 02$^{\rm{nd}}$ 2014} & {0.698 - 0.709} & {7.3}\\

		{\#71 12109} & {Jul 03$^{\rm{th}}$ 2014} & {0.349 - 0.379} & {6.8}\\
				\hline
		\end{tabular}	
	\label{Tab_A2}
\end{center}
\end{table}

\newpage

\footnotetext[1]{leading spot}
\footnotetext[2]{following spot}

\newpage
\clearpage

\section{Outliers in CLV curves}
\label{sec:app03}

\begin{figure}[htbp]
	\centering
		\includegraphics[width=\columnwidth]{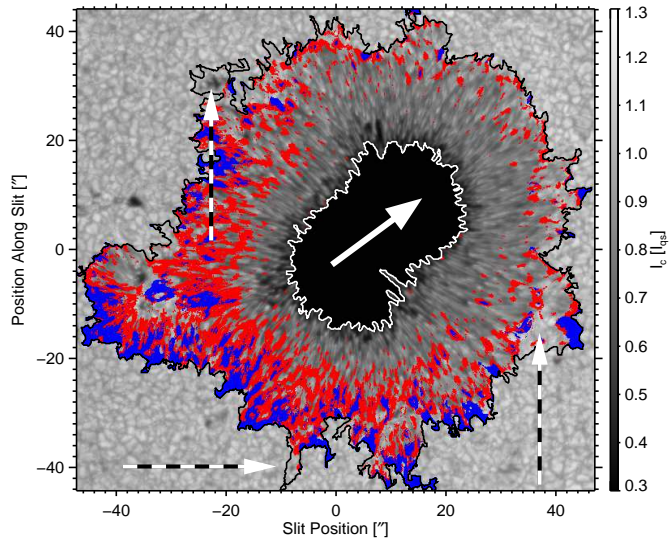}
		\caption{NOAA AR 10923 on November 13, 2007 with 3LPs (red) and RPPs (blue). The penumbra is asymmetric with a larger limb-ward side. It also includes regions where a more detailed inspection shows that the algorithm incorrectly identified granular regions (black and white arrows) as part of the penumbra. The white arrow points toward the center of the solar disk.}
		\label{fig:Franz_figA01}
\end{figure}

\begin{figure}[htbp]
	\centering
		\includegraphics[width=0.55\columnwidth]{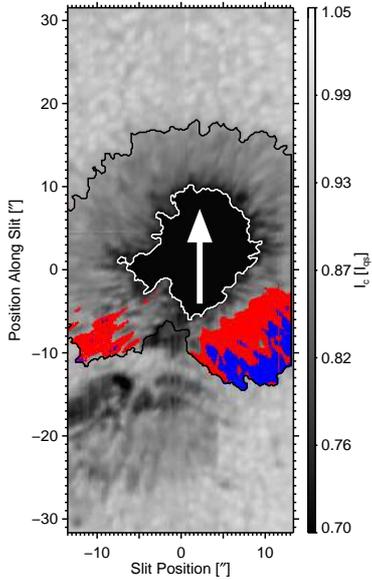}
		\caption{Leading spot of NOAA AR 12049 on April 30, 2014 was not scanned completely. The penumbra is asymmetric with a limb-ward part being detached from the sunspot. 3LPs and RPPs are indicated in red and blue. The white arrow points toward the center of the solar disk.}
		\label{fig:Franz_figA02}
\end{figure}

\begin{figure}[htbp]
	\centering
		\includegraphics[width=0.9\columnwidth]{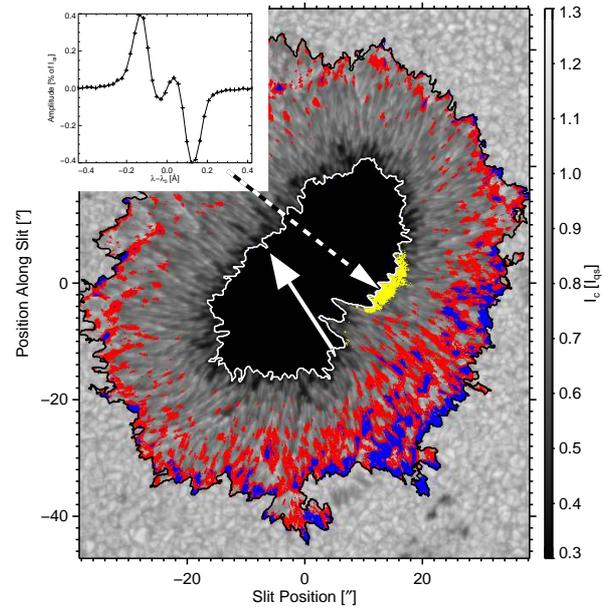}
		\caption{NOAA AR 10923 on November 14, 2007 with 3LPs (red) and RPPs (blue). Regular Stokes V profiles with a strong magneto-optical effect (inlay) that are incorrectly interpreted as 3LPs (yellow). The white arrow points toward the center of the solar disk.}
		\label{fig:Franz_figA03}
\end{figure}

\end{appendix}


\end{document}